\documentclass[useAMS,usenatbib]{mn2e}

\usepackage{verbatim,graphicx,epsfig,dcolumn,color}
\newcolumntype{.}{D{.}{.}{4}}
\newcolumntype{,}{D{.}{.}{2}}
\newcolumntype{;}{D{.}{.}{1}}
\newcommand{\nodata}{$\cdot\cdot\cdot$}
\newcommand{\lesssim}{{\lower-1.2pt\vbox{\hbox{\rlap{$<$}\lower5pt\vbox{\hbox{$\sim$}}}}}}
\newcommand{\gtrsim}{{\lower-1.2pt\vbox{\hbox{\rlap{$>$}\lower5pt\vbox{\hbox{$\sim$}}}}}}

\title[Pulsation-triggered winds among AGB stars]{Pulsation-triggered dust production by asymptotic giant branch stars}
\author[I. McDonald et al.]{I.~McDonald$^{1}$\thanks{E-mail: mcdonald@jb.man.ac.uk}, E.~De~Beck$^{2}$, A.~A.~Zijlstra$^{1,3}$, E.~Lagadec$^{4}$\\
$^{1}$Jodrell Bank Centre for Astrophysics, Alan Turing Building, Manchester, M13 9PL, UK\\
$^{2}$Department for Earth and Space Sciences, Chalmers University of Technology, Onsala Space Observatory, 43992, Onsala, Sweden\\
$^{3}$Department of Physics \& Laboratory for Space Research, Univerity of Hong Kong\\
$^{4}$Universit\'e ́C\^ote d{'}Azur, Observatoire de la C\^ote d{'}Azur, CNRS, Lagrange, France\\
}

\begin{document}

\date{Accepted 9999 December 32. Received 9999 December 32; in original form 9999 December 32}

\pagerange{\pageref{firstpage}--\pageref{lastpage}} \pubyear{9999}

\maketitle

\label{firstpage}

\begin{abstract}
Eleven nearby ($<$300 pc), short-period (50--130 days) asymptotic giant branch (AGB) stars were observed in the CO $J = (2-1)$ line. Detections were made towards objects that have evidence for dust production ($K_{\rm s}-[22] \gtrsim 0.55$ mag; AK Hya, V744 Cen, RU Crt, $\alpha$ Her). Stars below this limit were not detected (BQ Gem, $\epsilon$ Oct, NU Pav, II Hya, CL Hyi, ET Vir, SX Pav). $K_{\rm s}-[22]$ colour is found to trace mass-loss rate to well within an order of magnitude. This confirms existing results, indicating a factor of 100 increase in AGB-star mass-loss rates at a pulsation period of $\sim$60 days, similar to the known `superwind' trigger at $\sim$300 days. Between $\sim$60 and $\sim$300 days, an approximately constant mass-loss rate and wind velocity of $\sim$3.7 $\times$ 10$^{-7}$ M$_\odot$ yr$^{-1}$ and $\sim$8 km s$^{-1}$ is found. While this has not been corrected for observational biases, this rapid increase in mass-loss rate suggests a need to recalibrate the treatment of AGB mass loss in stellar evolution models. The comparative lack of correlation between mass-loss rate and luminosity (for $L \lesssim 6300$ L$_\odot$) suggests that the mass-loss rates of low-luminosity AGB-star winds are set predominantly by pulsations, not radiation pressure on dust, which sets only the outflow velocity. We predict that mass-loss rates from low-luminosity AGB stars, which exhibit optically thin winds, should be largely independent of metallicity, but may be strongly dependent on stellar mass.
\end{abstract}



\section{Introduction}
\label{IntroSect}

Stellar death among low- and intermediate-mass (0.8--8 M$_\odot$) stars is caused by catastrophic mass loss. In such stars, the mass-loss rate can greatly exceed the hydrogen nuclear-burning rate\footnote{The mass-to-light conversion ratio of the $^1$H $\rightarrow$ $^4$He reaction is 0.00717. Thus, the nuclear-burning rate of a star is $\sim 0.00717 L/c^2$, or $\sim$10$^{-7}$ M$_\odot$ yr$^{-1}$ for a 10\,000 L$_\odot$ star.}. However, the drivers of this mass loss are poorly known. Canonical theory dictates that magneto-acoustic effects (magnetic reconnection and/or Alfv\'en waves) support a fast ($\sim$30--300 km s$^{-1}$) but feeble ($\sim$10$^{-14}$--10$^{-8}$ M$_\odot$ yr$^{-1}$) wind throughout the star's main-sequence and giant-branch evolution \citep{DHA84,SC05,SC07,CS11}, possibly enhanced in later stages by bulk atmospheric motions (e.g.\ overshoot from convective cells into the upper atmosphere; cf.\ \citet{FH08}). Towards the end of the asymptotic giant branch (AGB) phase, the star becomes unstable to pulsations, which may levitate the outer atmosphere of the low-gravity star. The combination of moderately low temperatures and comparatively high pressures allows the condensation of dust around the star, and radiation pressure on this dust forces a wind from the star \citep[e.g.][]{Bowen88,HO18}.

However, calculations have shown that radiation pressure on conventional silicate dust is insufficient to drive a wind from oxygen-rich stars, and an additional mechanism is required \citep{Woitke06b}. Scattering of light by large ($\mu$m-sized) silicate grains has been proposed to drive the wind \citep{Hoefner08}. This has some observational confirmation \citep{NTI+12}, but forming large grains in low-mass-loss-rate stars, where a lower density of condensates should prevail, may be difficult \citep[e.g.][]{NBMG13}. Models of pulsation-enhanced, radiatively-driven stellar winds (both of individual stars and stellar populations) are now very sophisticated: they can convincingly reproduce the properties of well-established winds from AGB stars \citep[e.g.][]{BHAE15,DAGHS+17}. However they cannot reproduce the dusty but weaker winds seen in less-luminous ($\lesssim$5000 L$_\odot$), less-massive ($\lesssim$1 M$_\odot$) and/or metal-poor ([Fe/H] $\lesssim$ --0.5 dex) stars \citep[e.g.][]{BMB+15,BMS+15,MZW17}.

Observations have historically concentrated on the most luminous, most rapidly mass-losing stars: these bright AGB stars dominate the dust and gas return by intermediate-age populations to the local environment \citep{BSR+12}. Here, strong correlations between luminosity and both mass-loss rate and wind expansion velocity suggest that radiation pressure on dust is effective \citep{DTJ+15}. However, there have been few observations of less luminous stars ($L \lesssim 4000$ L$_\odot$).

The suggestion of a different link, between pulsations and the driving of stellar mass loss, was first suggested by \citet{WPF87}. It has been subsequently established that the development of an AGB superwind\footnote{The term ``superwind'' has received a variety of interpretations in the literature. Its origin broadly stems from a wind, with a mass-loss rate greatly exceeding that prescribed by \citet{Reimers75}, which ejects the remaining stellar envelope, allowing the formation of an optically visible planetary nebula \citep[e.g.][]{WC77,RV81,BW91}. We retain that definition here.} occurs at a period of $\sim$300 days \citep[e.g.][]{WC77,VW93}, with stars progressing quickly from optically thin winds to dust-enshrouded superwinds by periods of $\sim$700 days.

More recently, a similar mechanism has been described at shorter periods: stars with periods of $\gtrsim$60 days typically produce dust, while stars with $P \lesssim 60$ days typically don't. First described by \citet{ABC+01} and \citet{GSB+09} in Baade's Window, it was found to apply more generally to nearby stars by \citet{GvL07}. Largely forgotten in the literature, this phenomenon was examined again in \citet{MZ16}.

Circumstellar dust production is typically identified by infrared colour, e.g., $K_{\rm s}-[22]$ colour. Here, $K_{\rm s}$ represents a compromise between using a short wavelength where dust doesn't emit, and a long wavelength where dust doesn't absorb. The 22-$\mu$m flux represents a wavelength at which warm dust emits, but where stars remain relatively bright. Since both $K_{\rm s}$ and $[22]$ should fall on the star's Rayleigh--Jeans tail, $K_{\rm s}-[22] \approx 0$ implies a ``naked'' star, while $K_{\rm s}-[22] \gg 0$ indicates warm ($\gtrsim$300 K) dust in the line of sight. Thus, $K_{\rm s}-[22]$ and similar colours (e.g., $K-[12]$, $[3.6]-[24]$ or $[2.2]-[25]$) are commonly used as proxies for the dust-column opacity of a star \citep[e.g.][]{MvLS+11,RKJ+15}. However, translation of this to a physical mass-loss rate requires assumptions of the dust opacity per unit mass, the dust-condensation efficiency and the velocity structure of the wind.

CO both independently and more directly traces the wind properties \citep[e.g.][]{HO18}. The increase in $K_{\rm s}-[22]$ from 300 days is clearly linked to an increase in mass-loss rate, as derived from CO lines \citep[e.g.][]{DTJ+15}. However, the increase at 60 days has not been explored. It is unclear whether this represents a true increase in mass-loss rate, or simply a change in the properties of the stellar wind. To explore this, we examine literature gas (CO) mass-loss rates for stars around this boundary. The lack of suitable literature also led us to instigate a survey of nearby AGB stars in the CO $J = 2-1$ line, using the Atacama Pathfinder Experiment telescope (APEX). We show herein that the change at $P \sim 60$ days is a real increase in mass-loss rate, and represents a fundamental evolutionary change in how the mass-loss rate of these winds is set.

The remainder of our paper is set out as follows:
\begin{itemize}
\item Section \ref{LiteratureSourceSect} describes the literature data we have collated.
\item Section \ref{ObsSect} details our new APEX observations.
\item Section \ref{ResultsSect} merges the two samples.
\item Section \ref{DiscSect} discusses the results, highlighting both limitations in our data and future work that can be done to correct them.
\item Section \ref{ConcSect} reports our conclusions.
\end{itemize}


\section{Existing literature}
\label{LiteratureSourceSect}

\begin{figure}
\centerline{\includegraphics[height=0.45\textwidth,angle=-90]{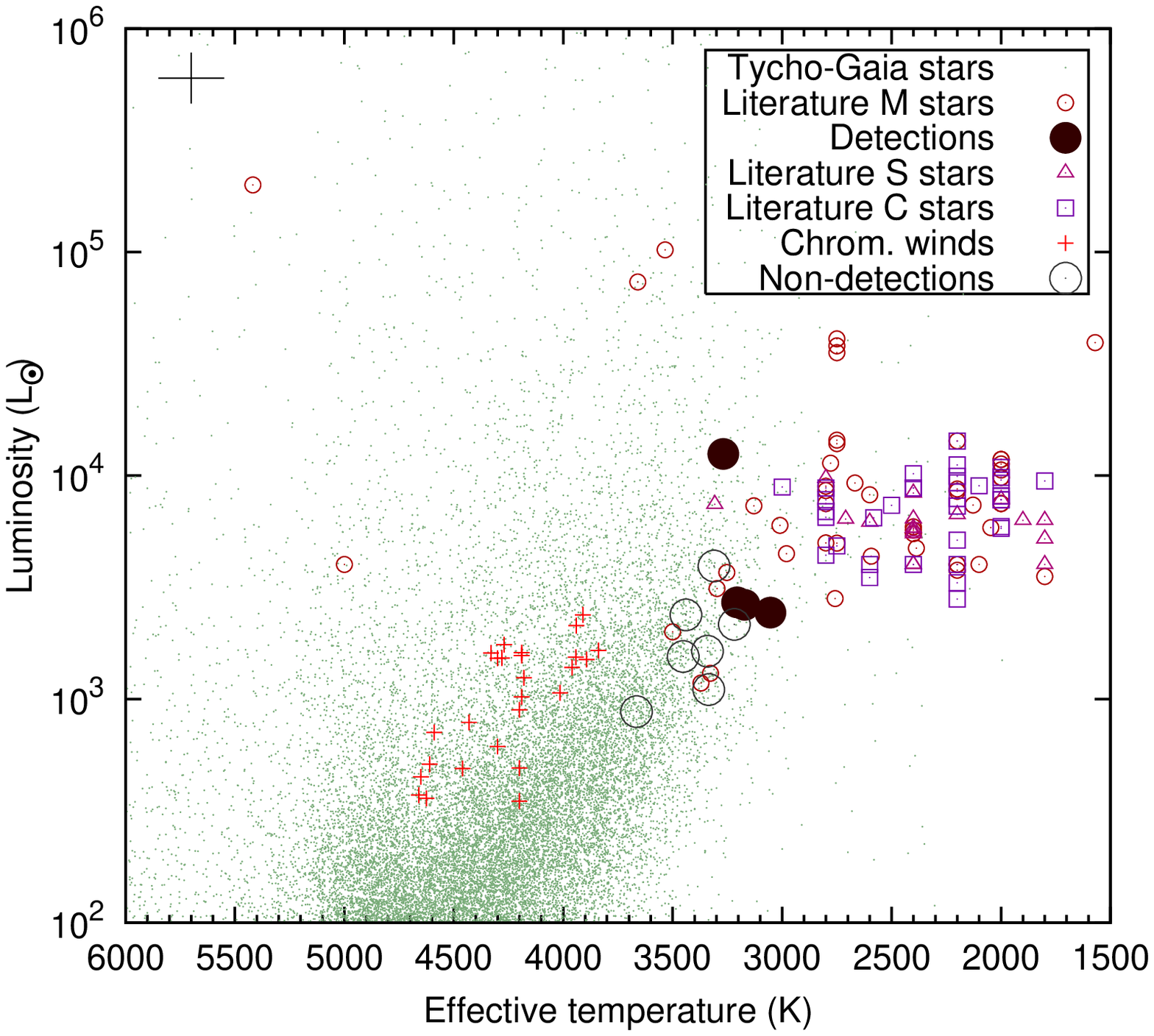}}
\centerline{\includegraphics[height=0.45\textwidth,angle=-90]{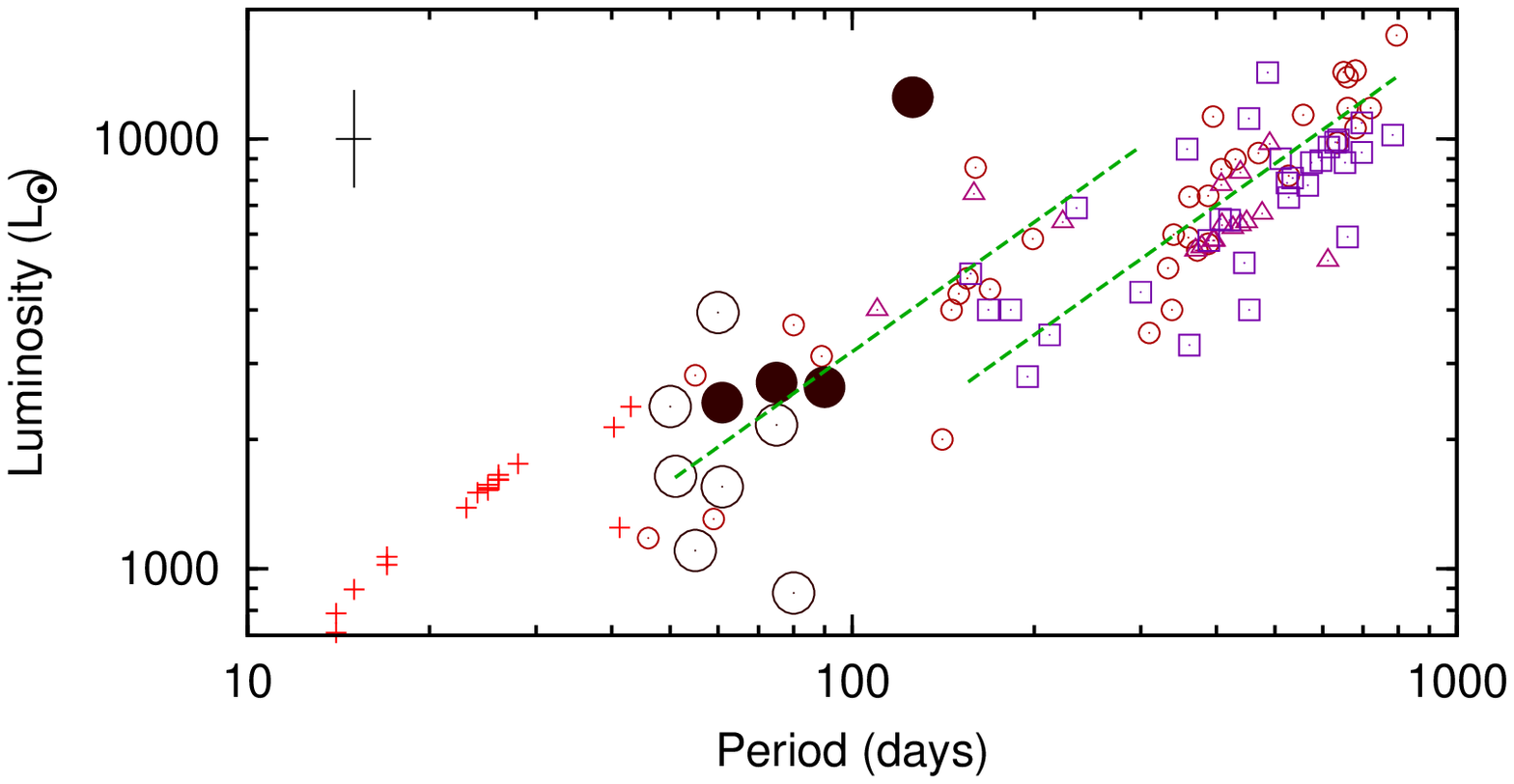}}
\caption{Hertzsprung--Russell diagram (top panel) of nearby stars. Stars from \emph{Gaia} Data Release 1 \citep{MZW17} are shown, alongside data from this paper and literature sources as indicated in the legend. Literature data are separated into M-, S- and C-type stars. New data are presented as both detections and non-detections. Stars with wind measurements from chromospheric indicators are also shown, as a representative uncertainties. The bottom panel shows the corresponding period--luminosity diagram, where the approximate loci of the fundamental (right) and first overtone (left) pulsations are shown as dashed lines. Note that some short-period pulsators are semi-regular variables, so may have ill-defined pulsation period.}
\label{HRDFig}
\end{figure}

\begin{figure}
\centerline{\includegraphics[height=0.45\textwidth,angle=-90]{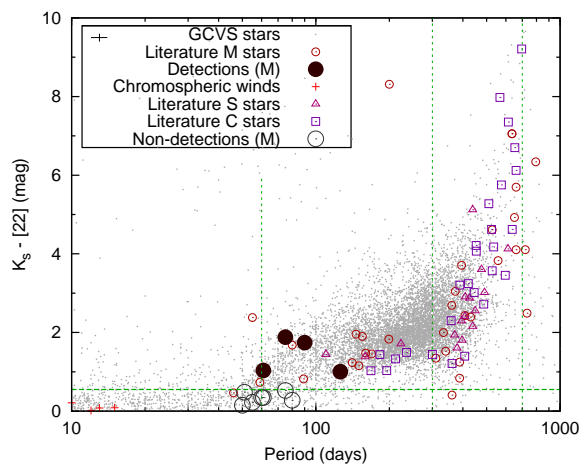}}
\caption{Period--infrared-excess diagram, as displayed in \citet{MZ16}. Symbols are as in Figure \ref{HRDFig}, or as described in the legend. The horizontal line denotes $K_{\rm s}-[22] = 0.55$ mag, set as a criterion for distinguishing stars with an without infrared excess. The vertical lines denote the major regime changes at $\sim$60, 300 and 700 days, discussed throughout the text.}
\label{PXSFig}
\end{figure}

\subsection{CO line obserations}
\label{LiteratureCOSect}

Collated literature observations of CO rotational lines exist in \citet{DBDdK+10}; \citet{SRO+13}; and \citet{DTJ+15}. For \citet{DBDdK+10}, we only use stars described as well-fit by a soft parabola. Additional CO observations of individual nearby AGB stars were extracted from \citet{HBCF94,WLBN+02,LLBGW08,LGT+10,LBMGL12,Groenewegen14,KHR+16,MZ16} and \citet{HRD+17}, forming our complete literature CO sample. Stellar parameters were also taken from these works. Due to the difficulties in measuring not only the distances of AGB stars (Section \ref{LiteratureGaiaSect}), but also measuring and defining their temperatures and radii (Section \ref{BiasSect}), we reduce our comparisons in the following to observational correlations with inferred luminosity and pulsation properties.

\subsection{Additional limits from chromospheric lines}
\label{LiteratureUncertainSect}

\begin{center}
\begin{table*}
\caption{Compiled literature data on chromospheric mass-loss rates.}
\label{RGBTable}
\begin{tabular}{lrrrrcl}
    \hline \hline
\multicolumn{1}{c}{Name} & \multicolumn{1}{c}{$P_{\rm adopted}$} & \multicolumn{1}{c}{$T_{\rm eff}$} & \multicolumn{1}{c}{$L$}       & \multicolumn{1}{c}{$\dot{M}$}             & \multicolumn{1}{c}{$v_{\rm exp}$}   & \multicolumn{1}{c}{References} \\
\multicolumn{1}{c}{\ }   & \multicolumn{1}{c}{(days)}            & \multicolumn{1}{c}{(K)}           & \multicolumn{1}{c}{L$_\odot$} & \multicolumn{1}{c}{(M$_\odot$ yr$^{-1}$)} & \multicolumn{1}{c}{(km s$^{-1}$)}   & \multicolumn{1}{c}{} \\
    \hline
$\omega$ Cen LEID 33011 &15& 4200 & 895 & 1.1 $\times 10^{-9}$ & 5.0  & VMC+11,MvLD+09 \\
$\omega$ Cen LEID 41039 &13& 4300 & 614 & 6.0 $\times 10^{-9}$ & 5.0  & VMC11,MvLD+09 \\
$\omega$ Cen LEID 37247 &12& 4200 & 492 & 3.2 $\times 10^{-9}$ & 15   & VMC11,MvLD+09 \\
$\omega$ Cen LEID 48321 &10& 4200 & 350 & 0.5 $\times 10^{-9}$ & 40   & VMC11,MvLD+09 \\
BD+17 3248       & \nodata & 4625 &  32 & \nodata              & 60   & DSS09 \\
HD 122563        & \nodata & 4625 & 360 & \nodata              & 140  & DSS09 \\
M13 IV-15        & 11 &      4650 & 449 & \nodata              & 30   & DSS09 \\
M13 L72          & 41 &      4180 &1247 & 2.8 $\times 10^{-9}$ & 11.0 & MAD09,Osborn00 \\
M13 L96          & 17 &      4190 &1023 & 4.8 $\times 10^{-9}$ & 19.0 & MAD09 \\
M13 L592         & 11 &      4460 & 489 & 2.6 $\times 10^{-9}$ & 8.5  & MAD09 \\
M13 L954         & 40 &      3940 &2133 & 3.1 $\times 10^{-9}$ & 12.0 & MAD09,Osborn00 \\
M13 L973         & 43 &      3910 &2382 & 1.6 $\times 10^{-9}$ & 6.5  & MAD09,Osborn00 \\
M15 K87          & 12 &      4610 & 511 & 1.4 $\times 10^{-9}$ & 9.0  & MAD09,MDS08 \\
M15 K341         & 25 &      4300 &1524 & 2.2 $\times 10^{-9}$ & 12.8 & MAD09,MDS08 \\
M15 K421         & 26 &      4330 &1611 & 1.9 $\times 10^{-9}$ & 10.0 & MAD09,MDS08 \\
M15 K479         & 28 &      4270 &1754 & 2.3 $\times 10^{-9}$ & 12.0 & MAD09,MDS08 \\
M15 K757         & 25 &      4190 &1567 & 1.3 $\times 10^{-9}$ & 9.5  & MAD09,MDS08,MvLDB10 \\
M15 K969         & 14 &      4590 & 710 & 1.4 $\times 10^{-9}$ & 7.5  & MAD09,MDS08 \\
M92 VII-18       & 26 &      4190 &1614 & 2.0 $\times 10^{-9}$ & 15.0 & MAD09 \\
M92 X-49         & 25 &      4280 &1528 & 1.9 $\times 10^{-9}$ & 15.0 & MAD09 \\
M92 XII-8        & 14 &      4430 & 787 & 2.0 $\times 10^{-9}$ & 11.0 & MAD09 \\
M92 XII-34       & 10 &      4660 & 372 & 1.2 $\times 10^{-9}$ & 8.0  & MAD09 \\
NGC 2808 37872   & 17 &      4015 &1067 & 1.1 $\times 10^{-9}$ & 15   & MCP06 \\
NGC 2808 47606   & 26 &      3839 &1652 & 0.1 $\times 10^{-9}$ & 15   & MCP06 \\
NGC 2808 48889   & 25 &      3943 &1542 & 3.8 $\times 10^{-9}$ & 53   & MCP06 \\
NGC 2808 51454   & 24 &      3893 &1503 & 7.0 $\times 10^{-9}$ & 10   & MCP06 \\
NGC 2808 51499   & 23 &      3960 &1387 & 1.2 $\times 10^{-9}$ & 18   & MCP06 \\
    \hline
\multicolumn{7}{p{0.75\textwidth}}{References: Osborn00 = \citet{Osborn00}, MCP06 = \citet{MCP06}, MDS08 = \citet{MDS08}, DSS09 = \citet{DSS09}, MAD09 = \citet{MAD09}, MvLD+09 = \citet{MvLD+09}, MvLDB10 = \citet{MvLDB10}, VMC+11 = \citet{VMC+11}. Pulsation data is also sourced from \citet{Clement97}.}\\
    \hline\\
\end{tabular}
\end{table*}
\end{center}

The collated mass-loss rates from CO line measurements do not adequately cover mass loss from lower-luminosity stars. Generally, the mass-loss rates from these stars are too low to efficiently detect with CO rotational transitions. However, these stars typically have active chromospheres. Proxies for their mass-loss rates and wind outflow velocities can be derived from chromospherically active lines \citep{DHA84}.

To enhance our comparisons, we add to our literature data chromospheric mass-loss rates, wind outflow velocities, and luminosities for stars in the globular clusters M13, M15, NGC 2808 and $\omega$ Cen (Table \ref{RGBTable}). These were sourced from \citet{MCP06,MDS08,MAD09,MvLD+09,VMC+11}, and from \citet{DSS09} for the field stars BD+17 3248 and HD 122563. $V$-band pulsation data for some sources in M13 and M15 have been sourced from \citet{Clement97} and \citet{MvLDB10}. Stars without known pulsation data have been assigned a pulsation period, as detailed below.

We must also add a note of caution here. Good estimates of mass loss from RGB stars are largely restricted to globular clusters and a few halo field stars. The shortfalls of this method are discussed in Section \ref{BiasSect}.

Most of the stars with chromospheric winds can be classified as non-variable ($\Delta V \ll 0.1$ mag), either by their absence in specific surveys \citep[references in][]{Clement97}, by \emph{Gaia} DR2, or by both. Three of the brightest stars have periods in the range of 40--42 days \citep{Osborn00}. For the remainder of the chromospheric detections in globular clusters, we assign periods based on the $b3$ sequence of small-amplitude variables from \citet[][their figure 9]{TSI+13}. It is possible some periods will be on the $b2$ sequence (or, rarely, the $b1$ sequence) at twice or thrice the periods quoted. However, $b3$ was chosen as it best fits the relation described by \citet[][equation 5]{TSI+13} for each star in question, assuming a typical mass of 0.65 M$_\odot$, because stars of similar luminosity in globular clusters and other dwarf galaxies show these periods (e.g.\ \citet{MZS+14} and the small-amplitude variables of \citet{LW05}), and because the $b3$ sequence is the most populous in this period range, giving a higher probability that stars are actually found here.

The combined sample of literature objects are shown in context on the Hertzsprung--Russell diagram (Figure \ref{HRDFig}) and period--infrared-excess diagram (Figure \ref{PXSFig}). Figure \ref{MLRVexpFig} shows how the derived mass-loss rates and expansion velocities we measure compare to results found in the literature.

\subsection{\emph{Gaia} DR2 and the uncertain distances to sources}
\label{LiteratureGaiaSect}

\begin{center}
\begin{table*}
\caption{Updated literature properties of stars, based on new parallaxes.}
\label{LitTable}
\begin{tabular}{lrrrll}
    \hline \hline
\multicolumn{1}{c}{Name} & \multicolumn{1}{c}{$d_{\rm adopted}$} & \multicolumn{1}{c}{$L$}       & \multicolumn{1}{c}{$\dot{M}$}             & \multicolumn{1}{c}{Original}  & \multicolumn{1}{c}{New} \\
\multicolumn{1}{c}{\ }   & \multicolumn{1}{c}{(pc)}              & \multicolumn{1}{c}{L$_\odot$} & \multicolumn{1}{c}{(M$_\odot$ yr$^{-1}$)} & \multicolumn{1}{c}{reference} & \multicolumn{1}{c}{reference} \\
    \hline
R Lep   	& 418 & 5149	& $8.1 \times 10^{-7}$	& \citet{DTJ+15}     & \emph{Gaia} DR2 \\
X TrA   	& 282 & 3314	& $1.2 \times 10^{-7}$	& \citet{DTJ+15}     & \emph{Gaia} DR2 \\
V CrB   	& 842 & 9467	& $5.9 \times 10^{-7}$	& \citet{DTJ+15}     & \emph{Gaia} DR2 \\
RV Aqr  	& 858 & 11152	& $3.8 \times 10^{-6}$	& \citet{DTJ+15}     & \emph{Gaia} DR2 \\
S Cas   	& 460 & 5210	& $1.8 \times 10^{-6}$	& \citet{DTJ+15}     & \emph{Gaia} DR2 \\
AFGL 292 	& 253 & 3774	& $1.3 \times 10^{-7}$	& \citet{DTJ+15}     & \emph{Gaia} DR2 \\
R Leo   	& 114 & 3537	& $8.5 \times 10^{-8}$	& \citet{DTJ+15}     & \emph{Gaia} DR2 \\
S CrB   	& 418 & 5897	& $2.5 \times 10^{-7}$	& \citet{DTJ+15}     & \citet{ZZR+17} \\
RR Aql  	& 633 & 11269	& $3.4 \times 10^{-6}$	& \citet{DTJ+15}     & \citet{ZZR+17} \\
W Hya   	&  98 & 7330	& $1.3 \times 10^{-7}$	& \citet{DBDdK+10}   & \citet{ZZR+17} \\
RX Boo  	& 127 & 5983	& $2.4 \times 10^{-7}$	& \citet{DBDdK+10}   & \emph{Gaia} DR2 \\
V438 Oph	& 387 & 4468	& $3.5 \times 10^{-8}$	& \citet{DBDdK+10}   & \emph{Gaia} DR2 \\
VY Leo  	& 114 & 1179	& $2.8 \times 10^{-9}$	& \citet{Groenewegen14}&\emph{Gaia} DR2 \\
EU Del  	& 112 & 1306	& $2.7 \times 10^{-8}$	& \citet{MZ16}       & \emph{Gaia} DR2 \\
RT Vir  	& 226 & 4741	& $1.2 \times 10^{-6}$	& \citet{SRO+13}     & \emph{Gaia} DR2 \\
EP Aqr  	& 119 & 2816	& $3.4 \times 10^{-7}$	& \citet{DBDdK+10}   & \emph{Gaia} DR2 \\
S Cep   	& 531 & 14254	& $2.3 \times 10^{-6}$	& \citet{SRO+13}     & \emph{Gaia} DR2 \\
W Pic   	& 665 & 7367	& $4.2 \times 10^{-7}$	& \citet{SRO+13}     & \emph{Gaia} DR2 \\
RZ Peg  	&1117 & 8354	& $6.1 \times 10^{-7}$	& \citet{SRO+13}     & \emph{Gaia} DR2 \\
R Cas   	& 176 & 8960	& $1.3 \times 10^{-6}$	& \citet{SRO+13}     & \citet{ZZR+17} \\
R Crt   	& 249 & 8581	& $1.2 \times 10^{-6}$	& \citet{SRO+13}     & \emph{Gaia} DR2 \\
    \hline\\
\end{tabular}
\end{table*}
\end{center}

{\it Gaia} Data Release 2 \citep{GaiaDR2} has allowed us to revise the distances to some of our targets. However, many targets do not have accurate distances in \emph{Gaia} DR2, for reasons explored in Section \ref{BiasSect}. Where we revised the distance from the literature estimate, we use a $d^2$ scaling to adjust both the luminosities and mass-loss rates. Where altered from the original publications, revised data are presented in Table \ref{LitTable}.

\subsubsection{Stars with non-zero \emph{Hipparcos} parallaxes}

Direct parallactic distances are available for some stars from the \emph{Hipparcos} \citep{Perryman97} and \emph{Gaia} \citep{GaiaMission} satellites. Significant tension exists for many AGB stars beetween the parallax from \emph{Hipparcos} \citep{vanLeeuwen07} and \emph{Gaia} Data Release 2 \citep{GaiaDR2}: reasons behind this are explored in Section \ref{BiasSect}. This appears to correlate with greater amplitude and significance of excess noise in the \emph{Gaia} five-parameter (position, parallax and proper motion) solutions. While these effects could be mitigated by using the more-accurate proper motions from the combined \emph{Hipparcos}--\emph{Gaia} dataset \citep{GaiaDR1} to stabilise the parallax solution, red stars were not included in this dataset.

The \emph{Hipparcos} and \emph{Gaia} parallaxes can be treated independently. Without an objective method of distinguishing between the two distances, we can adopt a variance-weighted average of the parallax, and invert it to obtain a distance, with an uncertainty defined by either the average of the positive and negative quadrature-summed parallax uncertainties (if the parallaxes agree within their combined uncertainty), or the standard deviation of the two parallaxes (if tension exists between them). This neglects the Lutz--Kelker bias \citep{LK73}, although this is not expected to be a significant in this subset of observations where, while noisy, distances are comparatively well determined (see discussion in the Appendix of \citet{MZW17}).

VY Leo, EU Del and EP Aqr have \emph{Gaia} parallaxes in agreement with the \emph{Hipparcos} parallaxes, and we have adopted the variance-weighted average of both. RX Lep, SW Vir and T Cet have conflicting parallaxes: in all three cases the \emph{Gaia} parallax is much smaller and uncertain, and we retain the \emph{Hipparcos} parallax. Similarly, we retain the distance for R Leo of 114 $\pm$ 14 pc based on \citet{vanLeeuwen07}, as this represents the combination of the parallaxes of \emph{Hipparcos} and \citet{Gatewood92}. No \emph{Gaia} parallax exists for $\alpha$ Ori, as it is too bright.

In some cases, maser parallaxes are also available, providing a third independent reference. RT Vir is in both \emph{Hipparcos} and \emph{Gaia} DR2, and has very discrepant parallaxes between the two ($\varpi=7.38 \pm 0.84$ and $2.05 \pm 0.29$ mas, respectively) but also has a maser-based distance from very-long baseline interferometry (VLBI) of $\varpi = 4.14 \pm 0.13$ mas \citep{ZZR+17}, which we consider more trustworthy. Similarly, the maser parallax for R Cas ($\varpi = 5.67 \pm 1.95$ mas) is close to the \emph{Gaia} parallax ($\varpi = 5.34 \pm 0.24$ mas) but is marginally discrepant from the \emph{Hipparcos} parallax ($\varpi = 7.95 \pm 1.02$ mas), hence we adopt the maser parallax. We also adopt the maser parallaxes quoted in \citep{ZZR+17} for S CrB, RR Aql, W Hya and RX Boo.

\subsubsection{Stars without \emph{Hipparcos} parallaxes}

These stars have no alternative direct measure of distance to compare to, hence we must devise criteria to determine whether we find the \emph{Gaia} DR2 parallax believable. The distances used in the literature sub-mm sources (Section \ref{LiteratureSourceSect}) mostly come from assuming stars are fundamental-mode pulsators, and that luminosity is proportional to period. In this work, we seek to disentangle luminosity from period, so it is important we use parallax distances where possible. This is doubly so because the period--luminosity relation (actually a period--infrared-colour relation) has a steep luminosity dependence and an intrinsic spread of $\Delta\log(P) \sim 0.05$ in width \citep[e.g.][]{ITM+04,Wood15}, leading to a bolometric luminosity uncertainty of $\gtrsim$25 per cent, once a bolometric correction is applied to the infrared flux.

We have chosen to adopt \emph{Gaia} DR2 distances where the following conditions are met:
\begin{itemize}
\item The fractional uncertainty in the parallax is less than 0.2.
\item The ratio of the excess astrometric noise per point to the parallax is less than 0.4.
\item The ratio of the combined excess astrometric noise and parallax uncertainty to the parallax itself is less than 0.33.
\item The fraction of along-scan points flagged as ``bad'' is less than 0.1.
\item They are within 40 per cent of the distance estimate used to calculate the mass-loss rate.
\end{itemize}
For the remainder of objects, we retain pulsation distances.


\section{New observations}
\label{ObsSect}

\subsection{Target selection}
\label{ObsTargSect}

In the entire literature sample, there are CO measurements of only eight AGB stars with periods of $P < 150$ days. Mass-loss rates for these stars therefore mainly rely on the chromospheric mass-loss rates. The eight stars include the interacting binaries L$_2$ Pup and EP Aqr \citep{LKP+15,NHW+15,KHR+16,HRD+18}, reducing the number of single stars (or non-interacting binaries) to six: Y Lyn, g Her, R Sct, VY Leo, EU Del and RX Lep. Of these, only VY Leo (HD 94705) and EU Del have luminosities below the RGB tip and periods (46 and 59 days, respectively) less than the $\sim$60-day transition.

VY Leo and EU Del have CO mass-loss rates ($\log \dot{M} \approx -8.5$ and --7.5, respectively; \citealt{Groenewegen14,MZS+16}) that are below those of the other four stars ($\log \dot{M} \approx -6.8$ to --6.2; \citealt{LLBGW08,DBDdK+10,DTJ+15}). This suggests that total mass-loss rate does increase at around 60 days, but is insufficient to be considered as strong evidence. Consequently, we began observations with APEX telescope to obtain data on more targets.

Our target selection began with the set of fundamental parameters for \emph{Hipparcos} stars from \citet{MZB12} and \citet{MZW17}\footnote{VizieR tables J/MNRAS/427/343/table2 and J/MNRAS/471/770/table2, respectively.}. Where stars are duplicated, results from the latter were taken. From this, we selected stars with $d < 300$ pc, $L > 680$ L$_\odot$ and $T_{\rm eff} < 5500$ K, to yield a set of 562 nearby, bright, evolved stars\footnote{$L = 680$ L$_\odot$ is the minimum luminosity at which dust production by evolved stars appears to occur \citep{MBvL+11,MZB12,MZW17}. This was defined by the star RU Crt, defined here as being considerably more luminous. Hence the actual minimum luminosity is expected to be greater.}. By adding the two local optically enshrouded AGB stars (IK Tau and CW Leo) we can obtain a complete sample of bright RGB and AGB stars within 300 pc.

The fundamental parameters of these stars are derived from spectral energy distribution fitting, which does not provide accurate parameters for heavily enshrouded or highly extincted stars. Improved measures of luminosity and temperature were for R Leo and W Hya were sourced from \citet{BZvdL+97}, from \citet{MMV+12} for R Scl, for \citet{MRKC12} for CW Leo, and \citet{MLSF08} for IK Tau. Updated parameters for EU Del were also taken from \citet{MZ16}.

Of these stars, 121 were selected that had pulsation periods in the General Catalogue of Variable Stars (GCVS; \citealt{SDZ+06}\footnote{VizieR table B/gcvs/gcvs\_cat.}), of which 47 have pulsation periods in the range $50 < P < 130$ days. Additional periods were sourced for $\alpha$ Her and CL Hyi from \citep{WHP06}. Of these 49 stars, 18 have declinations above +20$^\circ$ (FS Com, BQ Ori, RX Cnc, BD Peg, RS Cnc, SV Lyn, RW CVn, $\rho$ Per, V1070 Cyg, g Her, OP Her, TU Aur, Y Lyn, AF Cyg, TU CVn, X Her, TT Per and SS Cep), meaning they could not be observed from APEX, and three already have published observations (the aforementioned RX Lep, EP Aqr and EU Del).

This leaves 28 stars, where were submitted as a distance-ranked list of targets for APEX observation. Of these, 11 were actually observed, the remainder being $\theta$ Aps, V763 Cen, GK Vel, S Lep, SU Sgr, IO Hya, T Crt, ER Vir, Z Eri, AG Cet, V450 Aql, GZ Peg, RT Cnc, V1057 Ori, $\tau_4$ Ser, GK Com and RV Hya.

\subsection{Updated distances to sources from \emph{Gaia} Data Release 2}
\label{ObsDistSect}

As previously mentioned (Section \ref{LiteratureGaiaSect}) the \emph{Gaia} Data Release 2 parallaxes for variable stars exhibit considerable errors in excess of their stated formal uncertainties. In our observed sample, AK Hya, RU Crt, CL Hyi, II Hya, ET Vir and $\epsilon$ Oct have a $>$2$\sigma$ tension between the \emph{Gaia} and \emph{Hipparcos} parallaxes, reaching 5$\sigma$ for RU Crt. For all stars except RU Crt, we obtain our final distance by adopting a variance-weighted average of the \emph{Gaia} and \emph{Hipparcos} parallaxes.

RU Crt was highlighted by \citet{MZB12,MZW17} as being the lowest-luminosity giant star with infrared excess in the \emph{Hipparcos}-- and \emph{Tycho}--\emph{Gaia} Astrometric Solutions (HGAS/TGAS) of \emph{Gaia} Data Release 1 \citep{GaiaDR1}. Such a low luminosity with substantial dust production is unusual, and the new distance from \emph{Gaia} returns it to a more typical luminosity, so we adopt it. It remains unusual due to its position in the period versus $K_{\rm s}-[22]$ diagram (Section \ref{ResultsSect}).

Given the change in distance, the luminosity of these sources needs revised, which we do by simply scaling luminosities from \citet{MZB12} and \citet{MZW17} using the inverse square law, $L \propto d^2$, to account for their new distances and to provide the luminosities quoted in Table \ref{ObsTable}.

\subsection{Data \& reduction}
\label{ObsDataSect}

\begin{figure*}
\centerline{\includegraphics[width=0.60\textwidth,angle=-90]{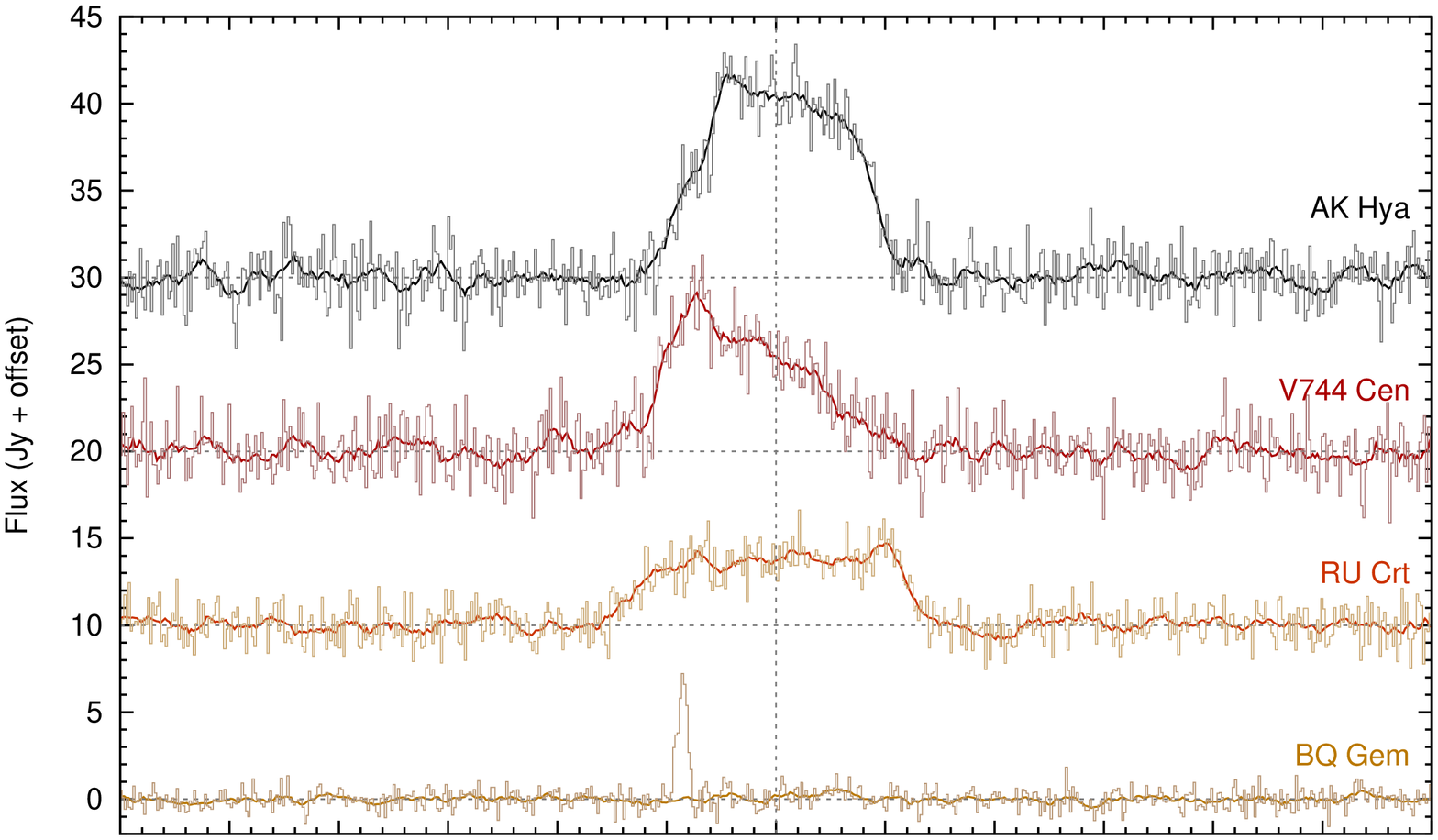}}
\centerline{\includegraphics[width=0.60\textwidth,angle=-90]{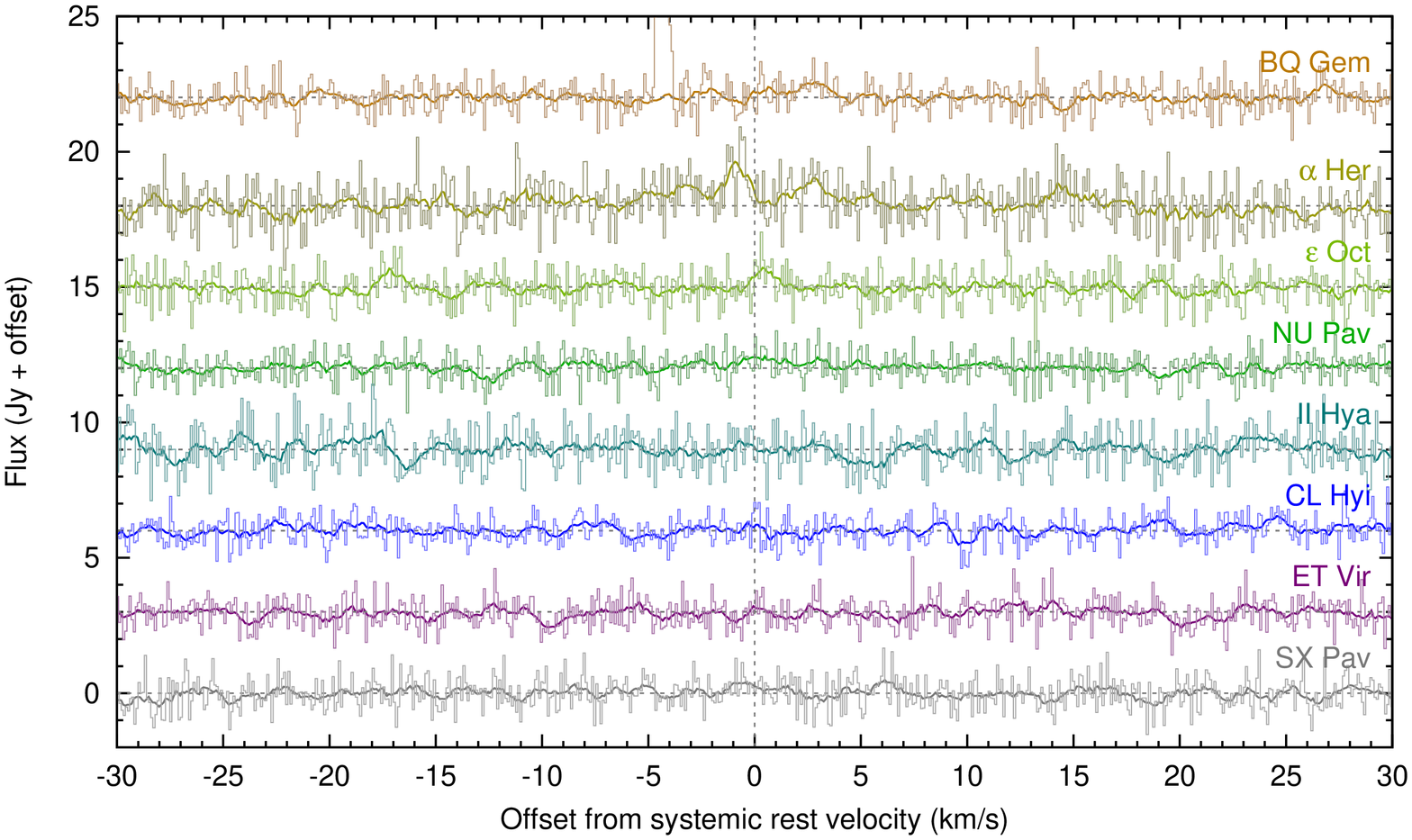}}
\caption{CO $J = 2 \rightarrow 1$ lines extracted from APEX spectra (points), shifted to the stellar velocity frame (except for CL Hyi, which has unknown velocity). The spike in the BQ Gem spectrum (reproduced in both panels) represents the masked interstellar component. The thicker lines have been spectrally binned by a boxcar of ten elements ($\sim$1 km s$^{-1}$) to show low signal-to-noise features. Table \ref{Obs2Table} lists the stellar $v_{\rm LSR}$ and noise measurements.}
\label{SpecFig}
\end{figure*}

\begin{center}
\begin{table*}
\caption{Literature properties of stars observed with \emph{APEX}. Distances from \emph{Hipparcos} \citep{vanLeeuwen07} and \emph{Gaia} \citep{GaiaDR2} are shown; asterisks ($^\ast$) denote stars with tension between the two results: see Section \ref{ObsDistSect} for details regarding the adopted distances. All stars are oxygen-rich.}
\label{ObsTable}
\begin{tabular}{lrrrrrrrrrl}
    \hline \hline
\multicolumn{1}{c}{Name} & \multicolumn{1}{c}{HIP} & \multicolumn{1}{c}{$d_{\rm Hip}$} & \multicolumn{1}{c}{$d_{\rm Gaia}$} & \multicolumn{1}{c}{$d_{\rm adopted}$} & \multicolumn{1}{c}{$P$} & \multicolumn{1}{c}{$K_{\rm s}-[22]$} & \multicolumn{1}{c}{$T_{\rm eff}$}  & \multicolumn{1}{c}{$L$} & \multicolumn{1}{c}{Status} \\
          \    & \     & \multicolumn{1}{c}{(pc)} & \multicolumn{1}{c}{(pc)} & \multicolumn{1}{c}{(pc)} & \multicolumn{1}{c}{(d)}  & \multicolumn{1}{c}{(mag)} & \multicolumn{1}{c}{$(K)$}   & \multicolumn{1}{c}{(L$_\odot$)} & \ \\
    \hline
        AK Hya & 42502 & 156\,$\pm$\,10 & 184\,$\pm$\,8  & 175\,$\pm$\,29 & 75 & 1.881 & 3206 &  2713 & Strong detection$^\ast$\\ 
      V744 Cen & 66666 & 158\,$\pm$\,8  & 165\,$\pm$\,10 & 161\,$\pm$\,6  & 90 & 1.743 & 3171 &  2545 & Strong detection\\ 
        RU Crt & 57800 & 132\,$\pm$\,17 & 239\,$\pm$\,12 & 239\,$\pm$\,12 & 61 & 1.034 & 3054 &  2436 & Strong detection$^\ast$\\ 
  $\alpha$ Her & 84345 & 110\,$\pm$\,16 & 101\,$\pm$\,5  & 104\,$\pm$\,6  &126 & 1.002 & 3269 & 12491 & Weak detection\\ 
        NU Pav & 98608 & 146\,$\pm$\,6  & 134\,$\pm$\,9  & 145\,$\pm$\,5  & 60 & 0.346 & 3311 &  3946 & No clear detection\\ 
        BQ Gem & 34909 & 166\,$\pm$\,21 & 172\,$\pm$\,8  & 174\,$\pm$\,10 & 50 & 0.149 & 3439 &  2385 & Interstellar line detected\\ 
        SX Pav &106044 & 126\,$\pm$\,5  & 133\,$\pm$\,5  & 132\,$\pm$\,8  & 51 & 0.480 & 3341 &  1640 & No clear detection\\ 
        CL Hyi & 11455 & 161\,$\pm$\,12 & 219\,$\pm$\,9  & 197\,$\pm$\,29 & 75 & 0.525 & 3218 &  2160 & No clear detection$^\ast$\\ 
        II Hya & 57613 & 167\,$\pm$\,8  & 129\,$\pm$\,6  & 149\,$\pm$\,24 & 61 & 0.345 & 3452 &  1552 & No clear detection$^\ast$\\ 
        ET Vir & 69269 & 141\,$\pm$\,6  & 163\,$\pm$\,7  & 155\,$\pm$\,22 & 80 & 0.276 & 3665 &   879 & No clear detection$^\ast$\\ 
$\epsilon$ Oct &110256 &  89\,$\pm$\,2  &  73\,$\pm$\,3  &  83\,$\pm$\,10 & 55 & 0.224 & 3335 &  1103 & No clear detection$^\ast$\\ 
    \hline\\
\end{tabular}
\end{table*}

\begin{table*}
\caption{Properties recovered from \emph{APEX} spectra. Radial velocities are from the CO lines, where they are detected (plain type), and from optical radial velocities (italic font; see text) when they are not. Otherwise, italics denote assumed values or, where uncertainties are quoted, very uncertain values. $I_{\rm CO,abs}$ represents the integrated flux of the CO line ($I_{\rm CO}$) scaled to a distance of 100 pc from the star, providing an absolute intensity scale.}
\label{Obs2Table}
\begin{tabular}{lrrrrrrrr}
    \hline \hline
\multicolumn{1}{c}{Name} & \multicolumn{1}{c}{Noise} & \multicolumn{1}{c}{$I_{\rm CO}$} & \multicolumn{1}{c}{$I_{\rm CO,abs}$}  & \multicolumn{1}{c}{$\beta$} & \multicolumn{1}{c}{$v_{\ast, {\rm LSR}}$} & \multicolumn{1}{c}{$v_{\rm exp}$} & \multicolumn{1}{c}{$\dot{M}$} & \multicolumn{1}{c}{$\chi^2_{\rm min}$} \\
          \    & \multicolumn{1}{c}{Jy channel$^{-1}$} & \multicolumn{1}{c}{(Jy km s$^{-1}$)}& \multicolumn{1}{c}{(at 100 pc)} & \  & \multicolumn{1}{c}{(km s$^{-1}$)}  & \multicolumn{1}{c}{(km s$^{-1}$)}  & \multicolumn{1}{c}{(M$_\odot$ yr$^{-1}$)} & \ \\
    \hline
        AK Hya & 1.388 &     88.5\,$\pm$\,1.2  &        293\,$\pm$\,97 &      0.644&             +19.4&       4.5&      1\,$\times$\,10$^{-6}$ &    1.67\\
      V744 Cen & 1.462 &     55.6\,$\pm$\,1.3  &        144\,$\pm$\,14 &{\it 0.413}&{\it       --15.0}&{\it  3.4}&$\sim$6\,$\times$\,10$^{-7}$ &    1.83\\
        RU Crt & 0.957 &     47.6\,$\pm$\,0.9  &        272\,$\pm$\,32 &      0.131&             +39.3&       5.9&      2\,$\times$\,10$^{-6}$ &    1.03\\
  $\alpha$ Her & 0.819 &      8.0\,$\pm$\,1.3  &       8.3\,$\pm$\,2.3 &{\it 0.696}&            --21.9&{\it 10.6}&$\sim$2\,$\times$\,10$^{-7}$ &    0.79\\
        NU Pav & 0.493 & {\it 2.2\,$\pm$\,0.6} &       4.6\,$\pm$\,1.6 &    \nodata&{\it         +3.5}&   \nodata&   $<$1\,$\times$\,10$^{-7}$ & \nodata\\
        BQ Gem & 0.474 & {\it 0.8\,$\pm$\,0.5} &       2.5\,$\pm$\,1.9 &    \nodata&{\it        --3.0}&   \nodata&   $<$1\,$\times$\,10$^{-7}$ & \nodata\\
        SX Pav & 0.570 & {\it 1.1\,$\pm$\,0.6} &       1.9\,$\pm$\,1.3 &    \nodata&{\it       --13.0}&   \nodata&   $<$8\,$\times$\,10$^{-8}$ & \nodata\\
        CL Hyi & 0.477 &               $<$2.6  &               $<$19.4 &    \nodata&           \nodata&   \nodata&   $<$2\,$\times$\,10$^{-7}$ & \nodata\\
        II Hya & 0.690 &               $<$1.1  &                $<$3.3 &    \nodata&{\it        --6.8}&   \nodata&   $<$7\,$\times$\,10$^{-8}$ & \nodata\\
        ET Vir & 0.506 &               $<$0.8  &                $<$2.6 &    \nodata&{\it       --18.6}&   \nodata&   $<$6\,$\times$\,10$^{-8}$ & \nodata\\
$\epsilon$ Oct & 0.640 &               $<$1.0  &                $<$0.9 &    \nodata&{\it        +31.0}&   \nodata&   $<$3\,$\times$\,10$^{-8}$ & \nodata\\
    \hline
\end{tabular}
\end{table*}
\end{center}

We obtained observations towards 11 of these stars (Table 1) using the Swedish Heterodyne Facility Instrument \citep[SHFI;][]{BLM+06,VML+08} on the APEX telescope \citep{GNS+06} between 2016 August 12 and November 28. The obtained spectra are centred on the CO $J=2-1$ rotational transition at 230.538 GHz and have a total spectral bandwidth of 4 GHz, corresponding to $\pm 2600$\,km\,s$^{-1}$ around the CO line, with a nominal resolution of $\approx$0.1\,km\,s$^{-1}$.

The delivered data are calibrated as antenna temperature, corrected for losses, $T_{\rm A}^*$. We reduced the data using the {\sc glidas/class} package\footnote{\texttt{http://www.iram.fr/IRAMFR/GILDAS}} by subtracting a first-degree polynomial from the spectrum, after emission-line masking. Conversion into physical units was done using the conversion factor\footnote{\texttt{http://www.apex-telescope.org/telescope/efficiency/}} $S_{\nu}/T_{\rm A}^*=39$\,Jy\,K$^{-1}$. Spectral channels at $|v_{\rm LSR}| > 100$ km s$^{-1}$ (compared to the rest frequency of the CO line) were assumed to have zero astrophysical flux. The noise per channel in the region of CO line was assumed to be equal to the standard deviation of fluxes in these high-velocity channels.

Radial velocities for AGB stars are not often known to better accuracy than a few km s$^{-1}$, as the pulsating atmospheres cause radial velocity variations and line doubling in their optical and near-IR spectra. Where available, we used published radial velocities from {\sc simbad}\footnote{http://simbad.u-strasbg.fr/simbad/} as a first estimate of the expected line centre. If no published velocity was available, a visual search of the spectrum was made to identify any emission peak.

\subsection{Final spectra}
\label{ObsSpecSect}

The final spectra for all stars are shown in Figure \ref{SpecFig}. There are three clear detections, of AK Hya, V744 Cen and RU Crt. A relatively square, symmetric profile indicates an optically thin, spherically symmetric wind around the star. While AK Hya and RU Crt show deviations from this, they better confirm to this classification than to optically thick or resolved winds, which would present as symmetric but more softened paraboloids. The spectrum of V744 Cen is more asymmetric, normally indicative of an asymmetric outflow. This leads to addition uncertainty in the parameters derived for this wind.

The spectrum of $\alpha$ Her shows a narrow peak (full-width half maximum [FWHM] $\sim$ 1 km s$^{-1}$) near the systemic velocity. This overlies a slight flux excess either side of the systemic velocity. This excess is significant ($\sim$6$\sigma$). While we cannot rule out an interstellar component for the narrow central line (the star lies close to, but not in, patchy far-infrared emission), it seems likely that we have made the first detection of CO around $\alpha$ Her. This would be significant as it was the original giant star in which mass loss was identified via optical circumstellar absorption lines \citep{Deutsch56}. The exact strength and shape of the CO emission cannot be accurately quantified from our data, but it seems to extend to $\sim\pm$5--10 km s$^{-1}$, commensurate with the $\sim\pm$10 km s$^{-1}$ indicated by \citet{Deutsch56}.

In BQ Gem, the primary peak is very narrow, and appears to be interstellar in origin\footnote{BQ Gem is projected onto a mid-infrared-bright cloud in Wide-field Infrared Survey (WISE; \citealt{CWC+13}) and other mid-infrared satellite images.}.  Consequently, we exclude the region containing the suggested interstellar line (between $v_{\rm LSR} = -1.9$ and --0.8 km s$^{-1}$) from further analysis of this star. There is no other convincing flux excess in this observation, so we consider BQ Gem to be a non-detection.

The remaining sources are not clearly detected, although some tentative detections are present. Statistically significant flux excesses exist at the systemic velocity for NU Pav ($\sim$3$\sigma$) and SX Pav ($\sim$2$\sigma$). The adopted stellar velocites and best-fit lines are listed in Table \ref{Obs2Table} for completeness, but treated as non-detections for the remainder of this paper as they cannot be accurately parameterised in terms of mass-loss rate or expansion velocity. A possible narrow component also exists in $\epsilon$ Oct, which we attribute to noise. II Hya, CL Hyi and ET Vir appear to have no emission to the limit of our observations.

From \citet{MZ16}, we expected that stars with $K_s-[22] \gtrsim 0.85$ mag would have well-established dusty winds, while those with $K_s-[22] \lesssim 0.85$ mag would not. Consequently, if mass-loss rate scales with $K_s-[22]$, we would expect those with $K_s-[22] \gtrsim 0.85$ mag to be detectable with our APEX observations out to a distance of a few hundred pc. Based on the $K$-[22] colours and \emph{Hipparcos} distances to these sources, detections were expected of AK Hya, V744 Cen, RU Crt and $\alpha$ Her, and non-detections were expected of the other sources \citep{MZ16}. Hence, with the exception of the interstellar line in BQ Gem and the faintness of $\alpha$ Her, the detections match our prior expectations.

\subsection{Determining $\dot{M}$ and $v_{\rm exp}$ from the observations}
\label{ObsParamSect}

The half-width of the CO line theoretically gives the wind expansion velocity, while the velocity-integrated line intensity is related to the CO mass-loss rate. Expansion velocities are easily calculated for lines representable by top-hat-like or parabolic functions, but become more difficult to define for clearly asymmetric lines, such as seen in V744 Cen.

Mass-loss rates from individual CO lines are necessarily approximate, as we do not precisely know the fractional abundance of CO, the radius where CO becomes dissociated by interstellar ultraviolet photons, the optical depth, the temperature structure, and the relative size of the envelope compared to the telescope beam (e.g., \citealt{MGH88,MZL+15,Groenewegen17}). However, mass-loss rates can be estimated from empirical relations, derived from stars where accurate mass-loss rates have been determined from multiple CO transitions. We here adopt the relation of \citet{DBDdK+10} (their equation 9) which, for unsaturated CO (2$\rightarrow$1) lines, states:
\begin{equation}
\dot{M} = \left(\frac{10^{-15} I_{\rm CO}}{\theta^{0.05} d^{-1.96} v_{\rm exp}^{-1.26} f_{\rm H,CO}^{0.79} T^{0.05} R^{0.24} R_{\rm in}^{0.10} \beta^{0.07}}\right)^{0.794}
\label{MdotEquation}
\end{equation}
where $I_{\rm CO}$ is the measured CO line intensity (in K km s$^{-1}$), $\theta$ is the telescope beam size (29$^{\prime\prime}$), $f_{\rm H,CO}$ is the adopted CO fraction relative to hydrogen (assumed to be 3 $\times$ 10$^{-4}$ following the original value listed for M-type stars\footnote{All 11 stars are classified as M-type in {\sc simbad}.} in \citet{DBDdK+10}), $R_{\rm in}$ is the dust-condensation radius of the wind in stellar radii ($R_\ast$) (assumed to be 3 $R_\ast$) and $\beta$ is a parameter describing the shape of the profile function as:
\begin{equation}
F_\nu(\nu) = F_{\nu,max} \left[ 1 - \left( \frac{v-v_\ast}{v_{\rm exp}} \right)^2 \right]^{\beta/2},
\label{BetaEquation}
\end{equation}
in the range $|v-v_\ast| \leq v_{\rm exp}$, for a line centred on $v_\ast$. Monte--Carlo sampling of the parameters was performed in the regime $v_\ast = [-10,10]$ km s$^{-1}$ from the literature value, $v_{\rm exp} = [1,20]$ km s$^{-1}$ and $\beta = [0.1,0.9]$. Where a line was not detected, we assumed a relatively large expansion velocity of 10 km s$^{-1}$, and cross-correlated this line (i.e., a boxcar function of 20 km s$^{-1}$ width) with the spectrum, identifying the strongest line within $\pm$10 km s$^{-1}$ of the stellar rest velocity. A mass-loss rate was calculated based on these parameters, and taken to be the upper limit to the mass-loss rate from the star. Note that this does not account for the order-of-magnitude systematic uncertainty in mass-loss rate that we apply generally below.

The primary assumptions that are required for Eq.~\ref{MdotEquation} to hold are: (1) a velocity profile similar to that of a radiation-driven wind; (2) a comparable incident UV flux; (3) optically thin winds with (at a given mass-loss rate) identical radial temperature and CO:H$_2$ profiles, which are (4) unresolved by the telescope's primary beam. The CO $J=2-1$ emission for stars of mass-loss rate $\sim$10$^{-7}$ is expected to be confined to a broad shell of a few $\times 10^{16}$ cm in radius \citep{THB+06}\footnote{Scalings are approximate: we estimate this value using a combination of their Figure 1 and Tables 3 \& 5, and scaling to a $\sim$5 km s$^{-1}$ wind.}. By this radius, radiation pressure and variations in stellar gravity are negligible, and variations associated with short-timescale events ($\lesssim$100 years) should be smoothed out, hence the velocity profile should be of negligible concern. Variation in interstellar UV flux is not well characterised but is not expected to depart wildly from that assumed by \citet{MGH88}, except in exceptional circumstances \citep{MZ15b}, suggesting the shell is dissociated at roughly the same radius. Dissociation should be accounted for in Eq.~\ref{MdotEquation}, as it is based on a set of observations which includes stars of similar mass-loss rates. However, it does mean that the expected shell size will be within the primary beam width, provided the star is $\gtrsim$50 pc away. The existing observations of EU Del and VY Leo indeed suggest this is the case: any resolution of the envelope or non-negligible optical thickness should also be observable as an inverted parabola (`two-horned') shape to the line profile, which is not observed. Given these assumptions, we expect Eq.~\ref{MdotEquation} to be accurate to a factor of a few in relative terms and within an order of magnitude in absolute terms.

The extracted mass-loss rates and expansion velocities are presented in Table \ref{Obs2Table}. Where undefined, we assume $\beta = 1$ and $v_{\rm exp} \leq 10$ km s$^{-1}$ in order to produce a limit to the mass-loss rate from our observations. Uncertainties are not listed on the stellar LSR velocity ($v_{\ast, {\rm LSR}}$), the wind expansion velocity ($v_{\rm exp}$), or the gas mass-loss rate ($\dot{M}$) as the formal errors described by the model are small, hence the (poorly constrained) systematic errors of the model dominate.

\section{Results}
\label{ResultsSect}

\begin{figure*}
\centerline{\includegraphics[height=0.90\textwidth,angle=-90]{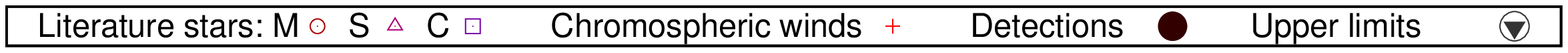}}
\centerline{\includegraphics[height=0.45\textwidth,angle=-90]{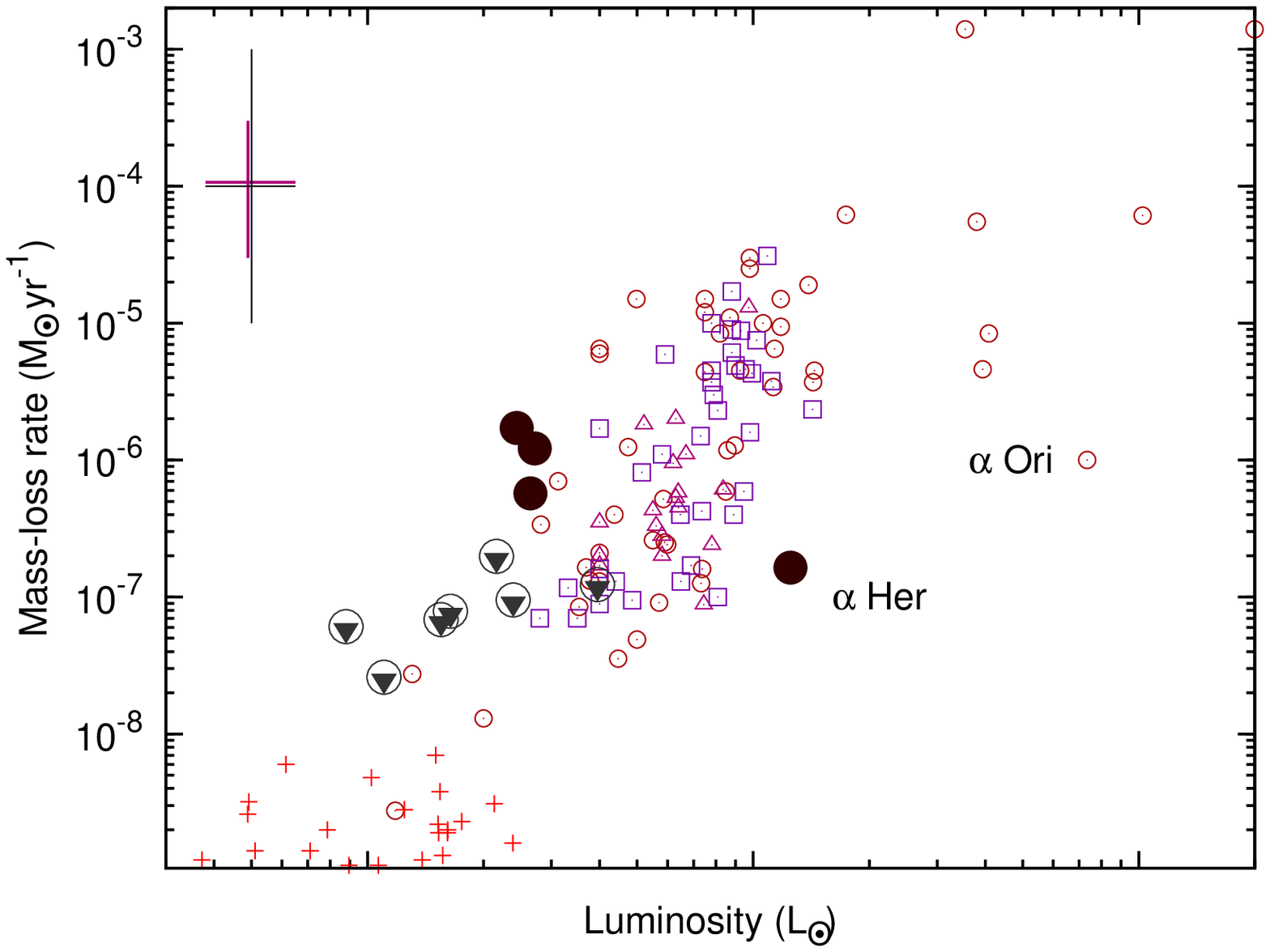}
            \includegraphics[height=0.45\textwidth,angle=-90]{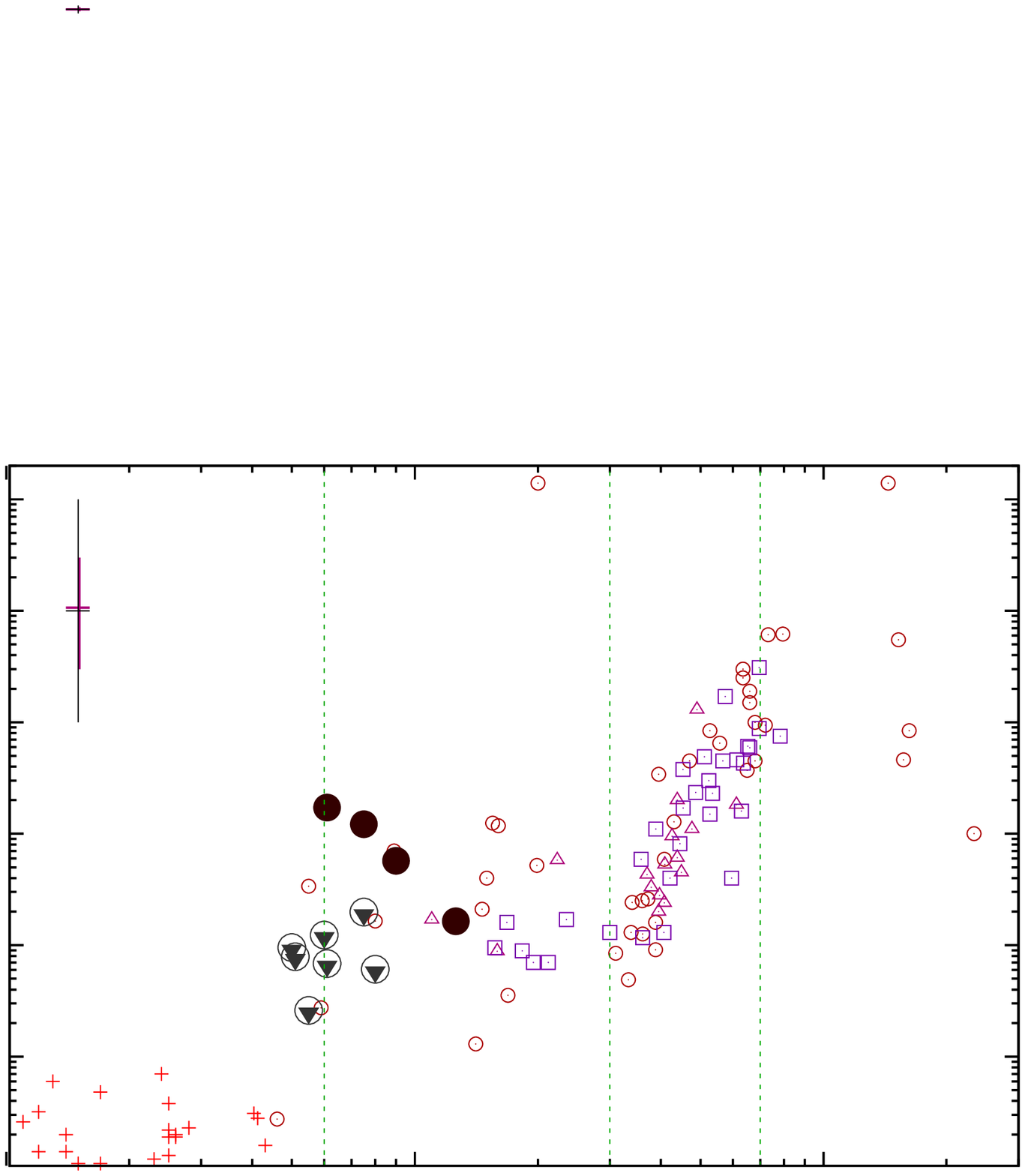}}
\centerline{\includegraphics[height=0.45\textwidth,angle=-90]{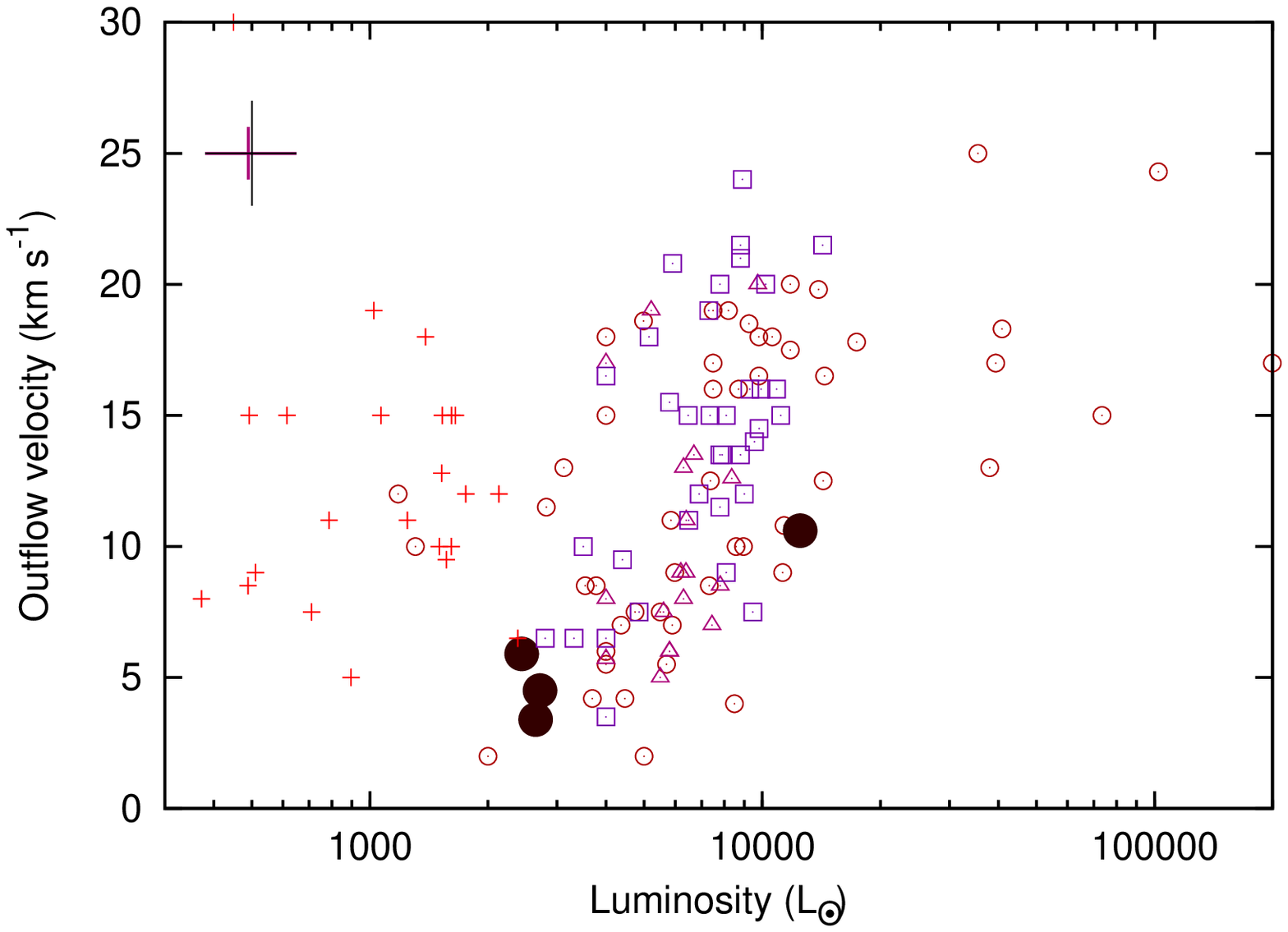}
            \includegraphics[height=0.45\textwidth,angle=-90]{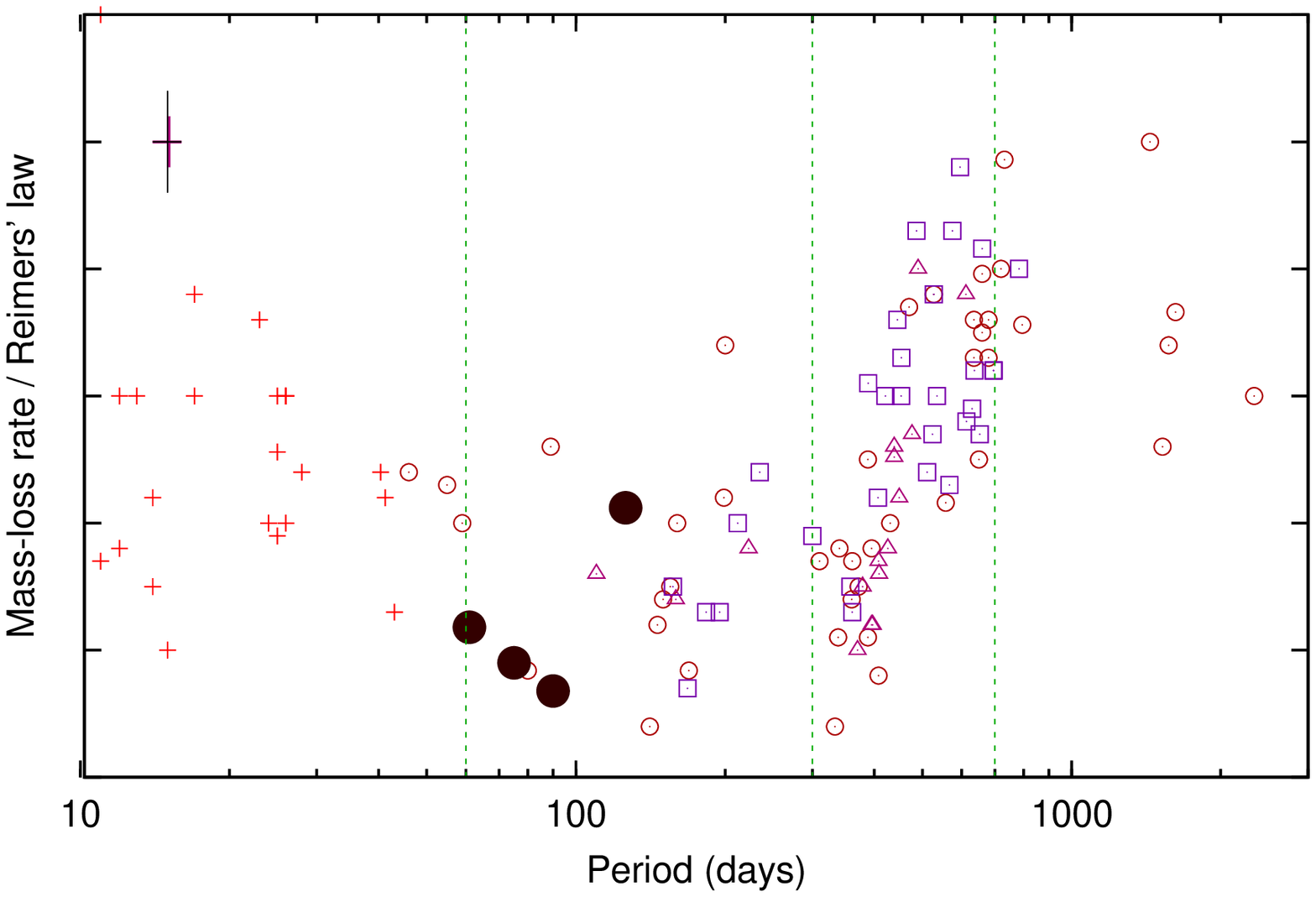}}
\caption{The relationship between total mass-loss rate and wind-expansion velocity, and pulsation period and luminosity, including data from literature sources. Symbols are as in Figure \ref{HRDFig} and in the legend, with upper limits on non-detections shown also as downward-pointing triangles. Representative errors are shown, showing {\it systematic} uncertainties in the literature (small, red lines) and our data (large, black lines). Relative uncertainties are likely to be considerably smaller.}
\label{MLRVexpFig}
\end{figure*}

\subsection{Observed trends}
\label{LiteratureObsSect}

\begin{figure}
\centerline{\includegraphics[height=0.45\textwidth,angle=-90]{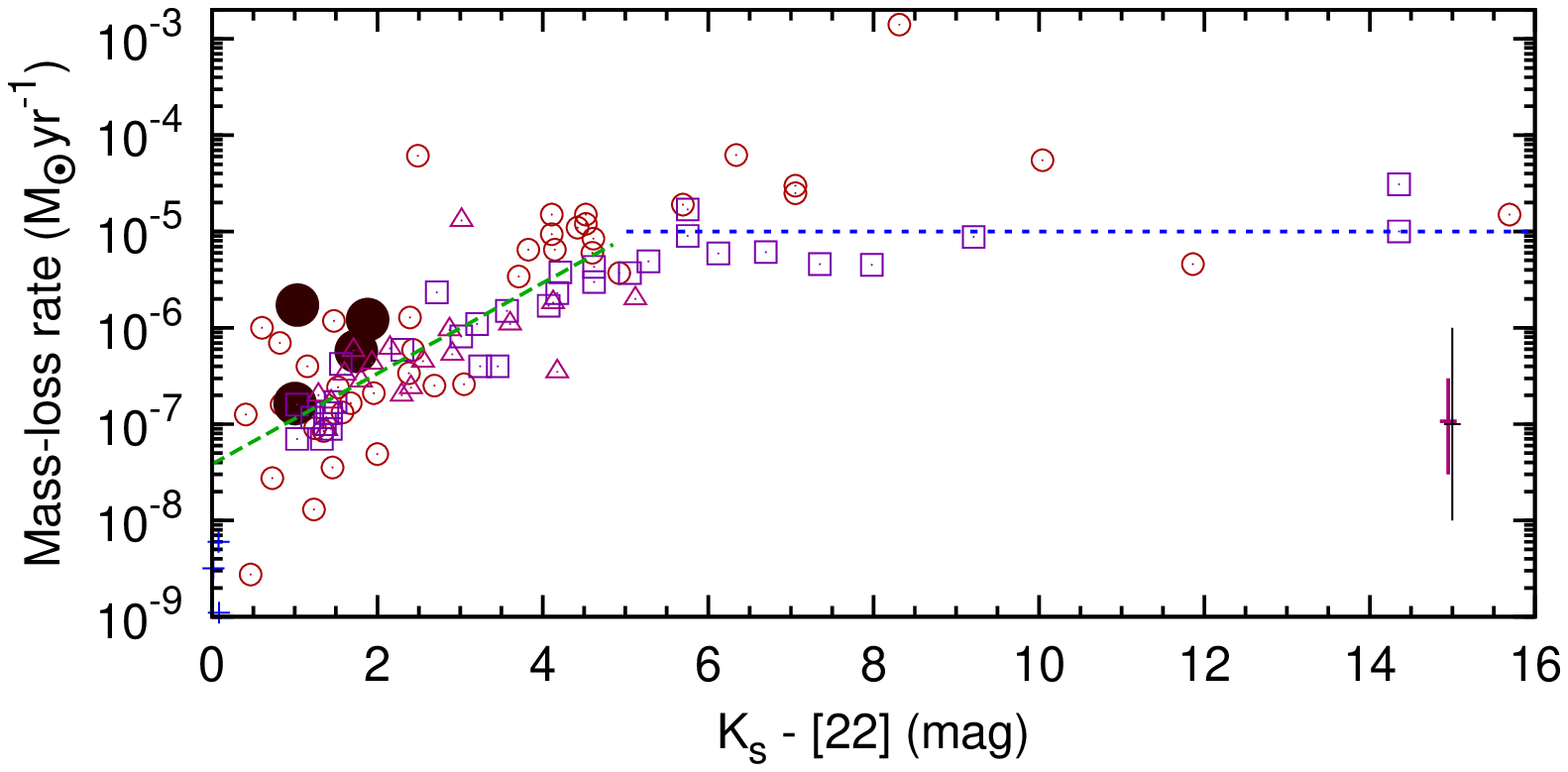}}
\centerline{\includegraphics[height=0.45\textwidth,angle=-90]{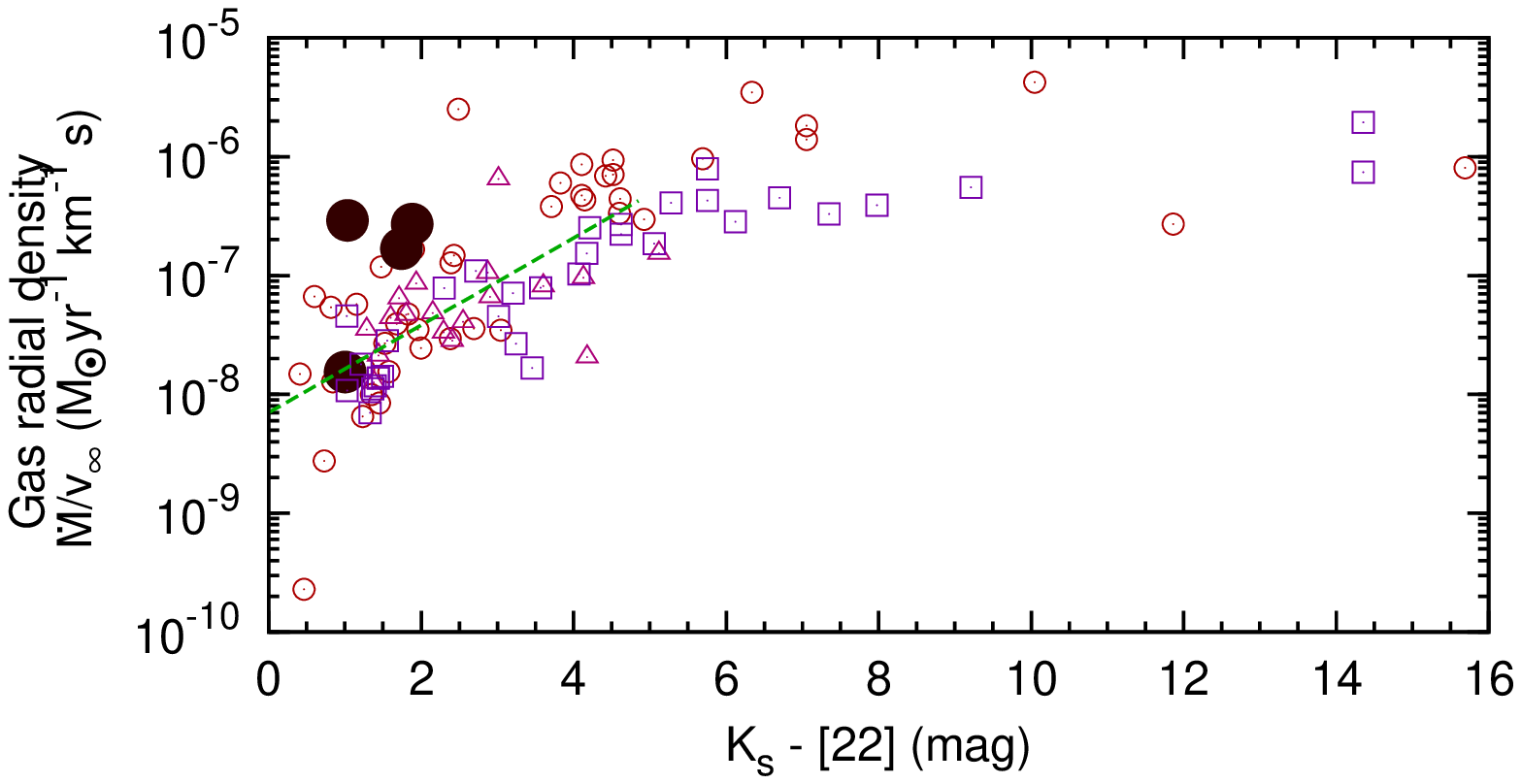}}
\caption{As Figure \ref{MLRVexpFig}, but showing the dust column opacity ($K_{\rm s}-[22]$ colour) versus (top panel) the gas mass-loss rate and (bottom panel) the gas column density. The green line is the best fit to the regime $K_s - [22] < 5$ mag, as listed in Section \ref{EmpiricalSect}. Representative systematic errors are shown; relative uncertainties are likely to be considerably smaller.}
\label{ColDensFig}
\end{figure}

Complex structures are seen in the combined dataset (Figure \ref{MLRVexpFig}) and in the mass-loss rate versus $K-[22]$ infrared colour diagram (Figure \ref{ColDensFig}). Between $K-[22]$ = 0 and 5 mag, there is good correspondence between the mass-loss rate and infrared colour (relations are defined in Section \ref{EmpiricalSect}): the scatter from the relation shown in Figure \ref{ColDensFig} corresponds roughly to the order of magnitude uncertainty in the individual measurements of mass-loss rate. Consequently, statements regarding the period--infrared-colour diagram (Figure \ref{PXSFig}) should translate closely to statements made about the period--mass-loss-rate diagram (lower-left panel of Figure \ref{MLRVexpFig}).

The most striking changes in the period--infrared-excess and period--mass-loss-rate diagrams occur near the following periods\footnote{The corresponding parameters of $L$, $R$ and $T_{\rm eff}$ are estimated based on typical quantities for such stars, extracted from \citet{MZB12}, \citet{MZ16} and \citet{MZW17}. These parameters, and the periods themselves, are meant to be indicative only, and should not be adopted as the transition properties of any given star.}:
\begin{enumerate}
\item $P \approx 60$ days (corresponding to $L \sim 1500$ L$_\odot$, $R \sim 100$ R$_\odot$ and $T_{\rm eff} \sim 3500$ K). This corresponds to the onset of dusty mass loss, as identified by \citet{MZ16}, where mass-loss rises from $\sim$a few $\times$ 10$^{-9}$ M$_\odot$ yr$^{-1}$ to $\sim$1.5 $\times$ 10$^{-7}$ M$_\odot$ yr$^{-1}$\,$^[$\footnote{Calculated from 18 stars with $100<P<250$ days, averaged as $\log\dot{M} = -6.84$ dex [M$_\odot$ yr$^{-1}$] with a standard deviation of 0.44 dex.}$]$.
\item $P \approx 300$ days (corresponding to $L \sim 4000$ L$_\odot$, $R \sim 350$ R$_\odot$ and $T_{\rm eff} \sim 2800$ K). This corresponds to the 300-day rise of strong mass loss, as identified by \citet{VW93}, where mass loss begins to rise from $\sim$1.5 $\times$ 10$^{-7}$ M$_\odot$ yr$^{-1}$.
\item $P \approx 700$ days (corresponding to $L \sim 10\,000$ L$_\odot$, $R \sim 1000$ R$_\odot$ but without a clear temperature definition). It is not clearly seen in our diagram, due to lack of reliable $K_{\rm s}$ photometry for extremely dust-enshrouded stars in 2MASS and lack of heavily obscured stars in the GCVS. This corresponds to the plateau of \citet{VW93}, where mass-loss rates stabilise around $\sim$10$^{-5}$ M$_\odot$ yr$^{-1}$.
\end{enumerate}

Corresponding velocity signatures can be seen in Figure \ref{MLRVexpFig}, yet these are more widely scattered. The variation of both $\dot{M}$ and $v_{\rm exp}$ with $P$ and $L$ are investigated further in Section \ref{PulsOnsetSect}.

\subsection{Empirical relations}
\label{EmpiricalSect}

Between $K_{\rm s}-[22]$ = 0 and 5 mag, a linear regression provides the following relations between $K_{\rm s}-[22]$ colour and gas radial density, mass-loss rate, wind momentum flux and wind kinetic-energy flux:
\begin{eqnarray}
\log \left(\dot{M} / v_\infty\right) &=& (0.363 \pm 0.045) (K_{\rm s} - [22]) \nonumber \\ && - (8.109 \pm 0.124), \nonumber \\
\log \left(\dot{M}\right) &=& (0.469 \pm 0.048) (K_{\rm s} - [22]) \nonumber \\ && - (7.383 \pm 0.131), \nonumber \\
\log \left(\dot{M} v_\infty\right) &=& (0.574 \pm 0.056) (K_{\rm s} - [22]) \nonumber \\ && - (6.650 \pm 0.155), {\rm \ and} \nonumber \\
\log \left(\frac{1}{2} \dot{M} v_\infty^2\right) &=& (0.679 \pm 0.069) (K_{\rm s} - [22]) \nonumber \\ && - (5.916 \pm 0.188),
\label{K22MdotEqns}
\end{eqnarray}
for $\dot{M}$ in M$_\odot$ yr$^{-1}$ and $v_\infty$ in km s$^{-1}$. Note that these relations reflect \emph{only} the statistical scatter in these relations, and do \emph{not} include systematic uncertainties in determining mass-loss rate. In theory, these could modify the intercept by up to one dex, and modify the slope by a small factor. However, systematic errors are unlikely to be so large without affecting calibrated properties, such as the initial--final mass relation, planetary nebula luminosity function, or dust-to-gas ratio beyond the limits prescribed by observation.

In Figure \ref{PXSFig}, we have drawn a line at $K_{\rm s} - [22] = 0.55$ mag, to visually separate detections (ours and others') from our non-detections (as described in Section \ref{ObsSpecSect} and Table \ref{ObsTable}). The relationships described in Eq. \ref{K22MdotEqns} imply stars with mass-loss rates $\gtrsim$8 $\times$ 10$^{-8}$ M$_\odot$ yr$^{-1}$ should produce infrared colours in excess of this value.

The expected upper limits to the mass-loss rates of observed sample (Table \ref{Obs2Table}) are close to this level, suggesting our observational limits approximate those in the literature data. The only two stars with CO detections below this line are W Hya and VY Leo. W Hya's \emph{WISE} [22] magnitude (--3.624) is erroneously low compared to its \emph{IRAS} [12] and [25] magnitudes (--5.429 and --5.619, respectively), suggesting it truly lies above the $K_{\rm s}-[22] = 0.55$ mag line. It has not been included when calculating the above relations. VY Leo shows only a very mild dust excess, and is only detectable in CO (respectively) due to its relative proximity and the depth of observation (VY Leo was detected at 70 mJy by \citet{Groenewegen14}: roughly half our typical noise level after 3 km s$^{-1}$ binning).

In Section \ref{ObsSpecSect}, we demonstrated that stars above and below $K_{\rm s}-[22] \gtrsim 0.55$ mag are, respectively, detectable and undetectable in modestly deep millimetric CO observations. While our observations lack sufficient accuracy to precisely define this boundary in terms of a mass-loss rate, in this Section, we have further shown that stars above this limit generally lose $\dot{M} \gtrsim 8 \times 10^{-8}$ M$_\odot$ yr$^{-1}$, while stars below this limit generally have mass-loss rates below this value. Since most stars undergo this colour transition near a period of $\sim$60 days, this argues that the onset of dust production found at $\sim$60 days by \citet{MZ16} corresponds to a real increase in mass-loss rate, and not simply an increase in dust-condensation efficiency.


\section{Discussion}
\label{DiscSect}

\subsection{A three-stage wind-driving mechanism}
\label{DiscDriveSect}
\label{DiscDriveExpectSect}

Disentangling cause and effect in AGB stars is very difficult, rendering it near-impossible to be conclusive about the driving mechanism for any AGB-star wind. Nevertheless, a substantial body of evidence can be built from these observations.

To explain these phenomena, we invoke three wind-driving processes: a magneto-acoustically driven wind, a pulsation-driven wind, and a radiatively driven superwind. In any given star, all three mechanisms will contribute, but one is likely to dominate. We invoke a change from a magneto-acoustically dominated wind to a pulsation-driven wind at $P \approx 60$ days, and a transition to a pulsation-enhanced but dust-driven wind after $P \approx 300$ days. The maximum mass-loss rate achieved at $P \approx 700$ days may be linked to stars becoming optically thick, and transferring all their photon momentum into driving the wind.

\subsubsection{The magneto-acoustic wind and Reimers' law}
\label{ReimersSect}

\begin{figure}
\centerline{\includegraphics[height=0.45\textwidth,angle=-90]{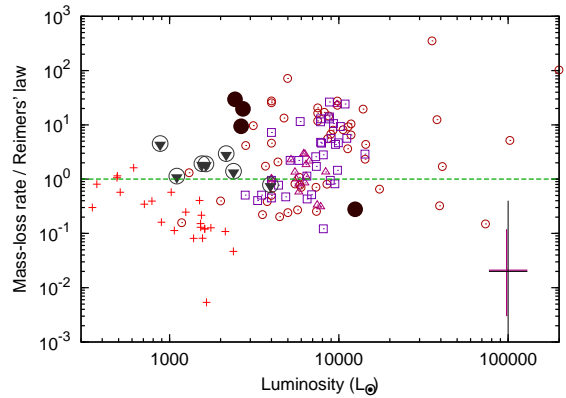}}
\caption{As Figure \ref{MLRVexpFig}, but showing the mass-loss rate divided by the expected value from \citet{Reimers75}, assuming $\dot{M} = 4 \times 10^{-13} \eta LR/M$, in solar units, where $\eta \approx 0.477$ \citep{MZ15b} and $M$ is assumed to be 1 M$_\odot$ (see text). The horizontal line denotes a correct prediction: values above indicate stars are losing mass faster than Reimers' law predicts. Representative systematic errors are shown; relative uncertainties are likely to be considerably smaller.}
\label{ReimersFig}
\end{figure}

Lower-luminosity stars are thought to be supported by magneto-acoustic winds, as strong pulsation and dust-production first appears in stars close to the RGB tip \citep[e.g.][]{MZB12}. However, differences in the masses of AGB and RGB stars have been found through direct \citep{MJZ11} and indirect methods \citep{MZ15b}. The evolutionary time over which this mass is lost ($\sim$0.2 M$_\odot$ over 200 Myr) indicates that mass-loss rates of at least $\dot{M} \gtrsim 10^{-9}$ M$_\odot$ yr$^{-1}$ must take place on both the RGB and AGB (see also \citet{LW05}). Mass loss during this stage is expected to take place via a magneto-acoustically heated chromosphere, as evidence by blue-shifted, chromospherically active lines seen in giant-branch stars \citep[e.g][]{DHA84}.

Historically, \citet{Reimers75} has been used to approximate mass loss from giant stars (see also \citealt{SC05,CS11}). Stellar evolution models still use it as the default choice of mass-loss law for magneto-acoustic winds, with its use now normally restricted to the wind of less-evolved, dustless stars, before some form of ``superwind'' in initiated \citep[e.g.][]{BMG+12,PCA+13}. Consequently, departures of stellar mass-loss rates from Reimers' law directly impact how the community models late-stage stellar evolution.

The masses of individual stars in our observations are unknown, except in the case of the globular cluster stars, which are constrained to be 0.53--0.80 M$_\odot$ for RGB stars \citep[e.g.][]{KSDR+09,GCB+10,MJZ11,MZ15b}. We assume these stars have $M = 0.65$ M$_\odot$.

We can estimate masses for field stars, based on a a typical initial mass function ($N\,dM \propto M^{-1.35}$), between 0.8 M$_\odot$ (the lowest mass star evolving through the AGB) and 8 M$_\odot$ (the highest mass AGB star). The central 68 per cent of stars will have initial masses between 1.0 and 4.6 M$_\odot$, and final masses between 0.5 and 1.0 M$_\odot$. A typical mass-losing star will be halfway between the two. Therefore, rounding down to account for a declining star-formation history in our Galaxy, we can expect a typical nearby AGB star to be $\sim$0.8 to 2.4 M$_\odot$ at present, with a median value just above 1 M$_\odot$. Hence, $M = 1 M_\odot$ has been assumed as a typical average present-day mass, with a conservative factor-of-two uncertainty to account for local variations and observing biases.

The ratio of observed mass-loss rates to Reimers' law is shown in Figure \ref{ReimersFig}. Except for the most extreme deviations, the large systematic uncertainties in mass-loss rates make it difficult to identify firm conclusions about the applicability of Reimers' law in general. Relative mass-loss rates should be more reliable, as internal uncertainties are expected to be several factors smaller (see Section \ref{ObsParamSect}).

Mass-loss rates from chromospheric lines become progressively sub-Reimers by the RGB tip ($\sim$2200 L$_\odot$), as are CO observations of the putative RGB star VY Leo \citep{Groenewegen14}, although EU Del \citep{MZ16} also falls into this category and has an approximately Reimerian mass-loss rate. Statistically, the ratio of log(observed/Reimers') mass-loss rate rises from --1.19 (standard deviation 0.71) for stars with $L = 1000-2000$ L$_\odot$ to +0.83 (0.63) for stars with $L = 2400-3400$ L$_\odot$. Even the large systematic uncertainties and observational biases cannot account for this factor of 100 difference, implying that mass-loss prescriptions on both the upper RGB and early AGB may need to be rethought.

\subsubsection{The onset of dusty mass loss set by pulsations}
\label{PulsOnsetSect}


\begin{figure}
\centerline{\includegraphics[height=0.45\textwidth,angle=-90]{legend.eps}}
\centerline{\includegraphics[height=0.45\textwidth,angle=-90]{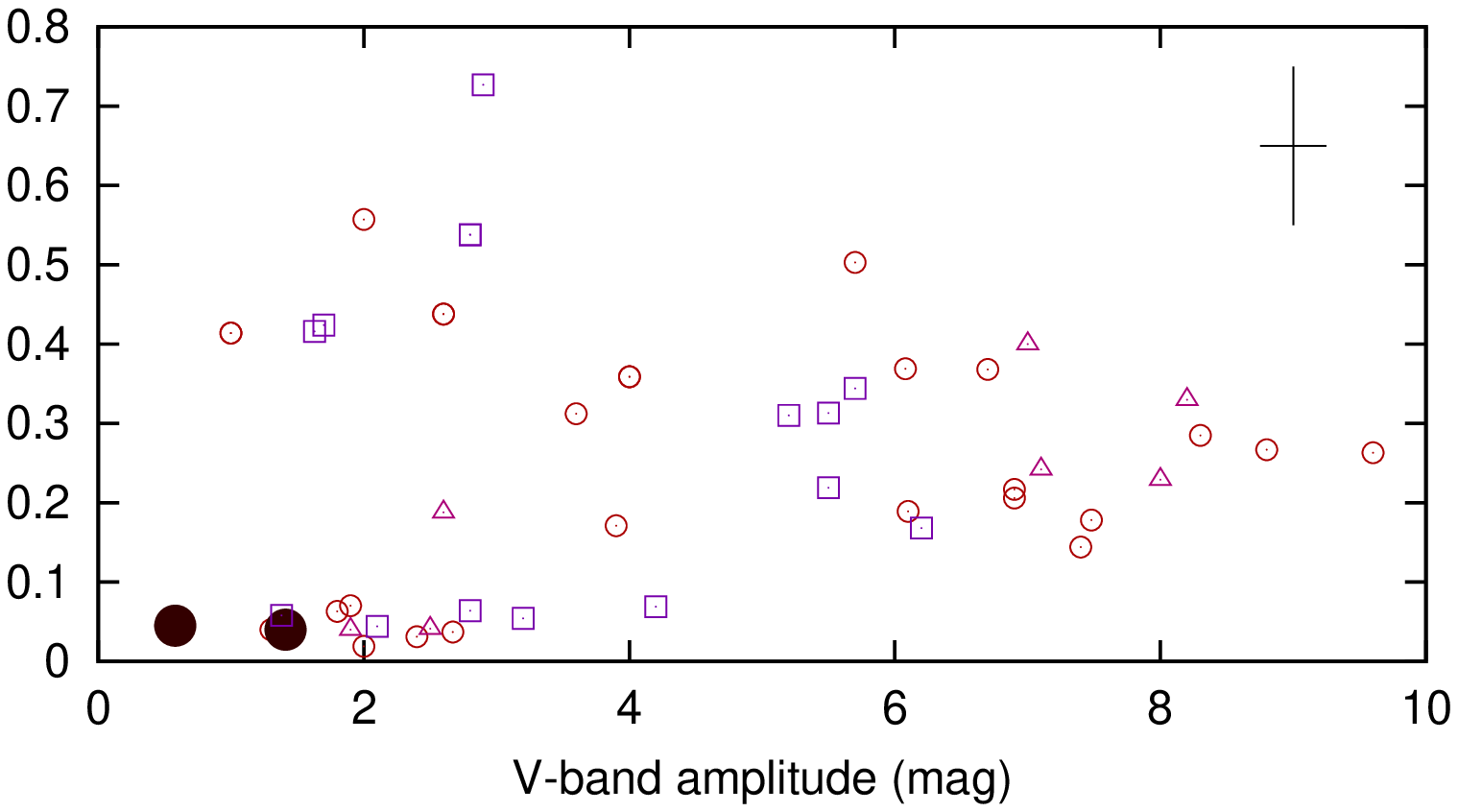}}
\centerline{\includegraphics[height=0.45\textwidth,angle=-90]{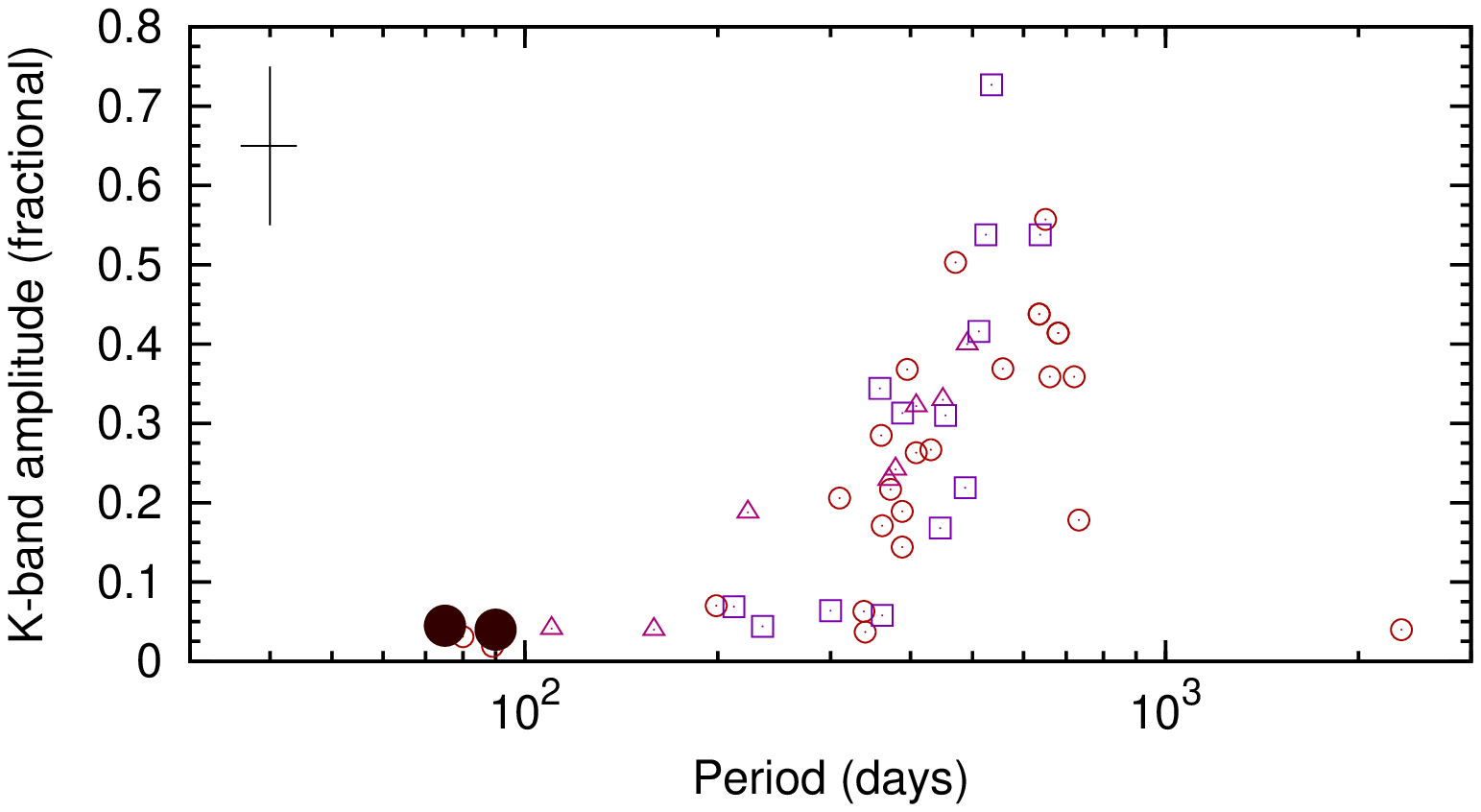}}
\centerline{\includegraphics[height=0.45\textwidth,angle=-90]{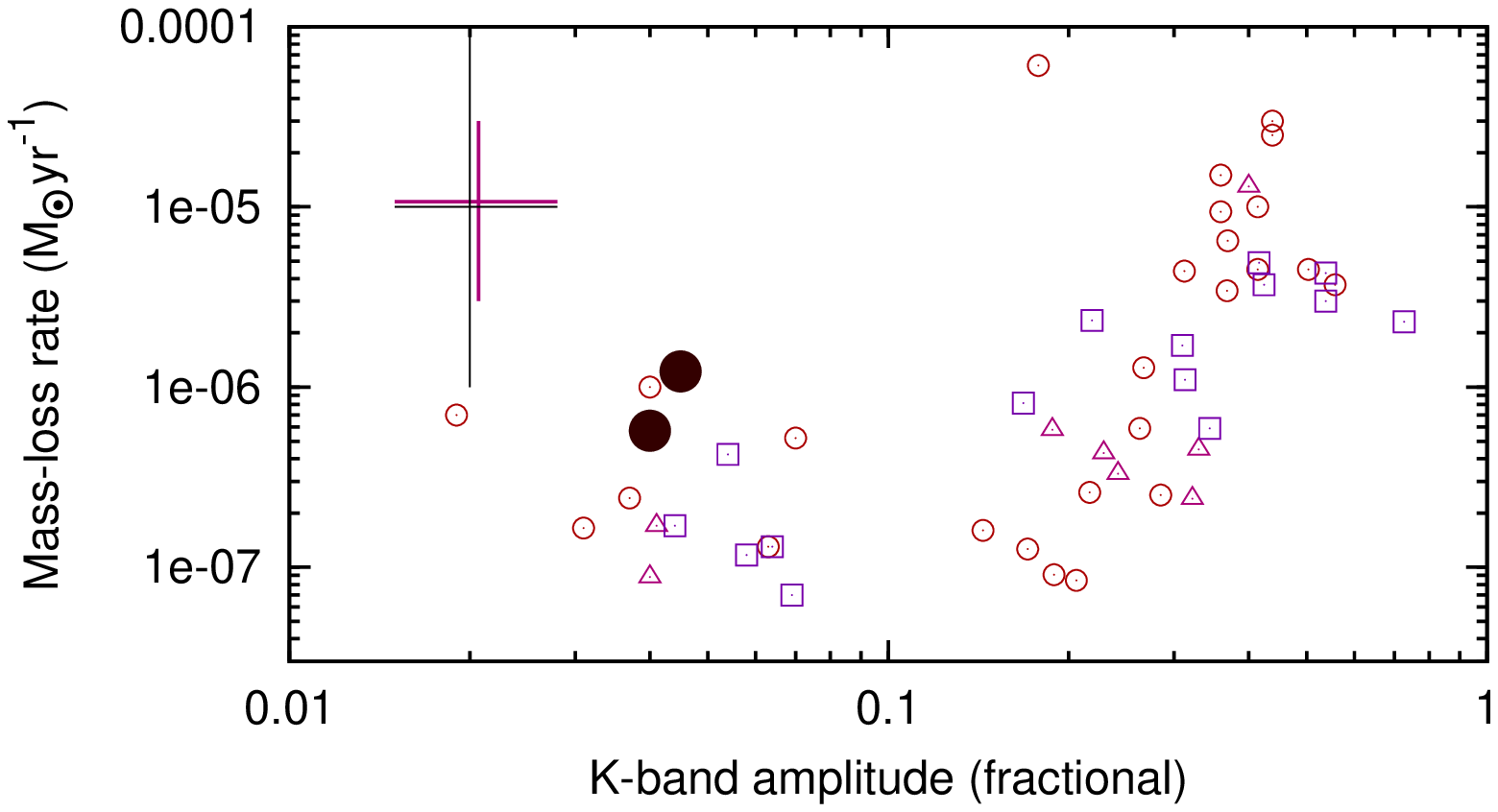}}
\caption{The $K_{\rm s}$-band (radial pulsation) amplitude versus $V$-band amplitude (top panel), pulsation period (middle panel) and stellar mass-loss rate (bottom panel). Symbols are as Figure \ref{MLRVexpFig}. Representative systematic errors are shown; relative uncertainties are likely to be considerably smaller.}
\label{KAmpMLRFig}
\end{figure}

\begin{center}
\begin{table}
\caption{Binned mass-loss rates for various period and luminosity combinations. See text for the reasons behind the choice of bins.}
\label{BinTable}
\begin{tabular}{@{}l@{}c@{}c@{}c@{}c@{}c@{}}
    \hline \hline
\multicolumn{1}{c}{Parameter} & \multicolumn{1}{c}{Mean}               & \multicolumn{1}{c}{St.}                   & \multicolumn{1}{c}{Mean}          & \multicolumn{1}{c}{St.}            & \multicolumn{1}{c}{No. of}   \\
\multicolumn{1}{c}{range}     & \multicolumn{1}{c}{$\log{\dot{M}}$}    & \multicolumn{1}{c}{dev.}                  & \multicolumn{1}{c}{$v_{\rm exp}$} & \multicolumn{1}{c}{dev.}           & \multicolumn{1}{c}{stars} \\
    \hline
$40 < P \leq 46$ days       & --8.59 & 0.13 & 10.4 &  2.6 & 4\\
$55 \leq P < 65$ days       & --6.60 & 0.88 &  9.1 &  2.9 & 3\\
$65 \leq P < 100$ days      & --6.27 & 0.37 &  6.3 &  4.5 & 4\\
$100 \leq P < 150$ days     & --7.03 & 0.57 &  6.7 &  3.6 & 4\\
$150 \leq P < 200$ days     & --6.78 & 0.42 &  7.1 &  2.3 & 10\\
$200 \leq P \leq 235$ days  & --6.72 & 0.46 & 10.3 &  1.3 & 4$^\ast$\\
$300 \leq P < 400$ days     & --6.64 & 0.40 &  7.7 &  3.0 & 18\\
$400 \leq P < 500$ days     & --6.04 & 0.48 & 13.2 &  4.7 & 17\\
$500 \leq P < 600$ days     & --5.45 & 0.47 & 16.3 &  4.8 & 9\\
$600 \leq P < 700$ days     & --5.10 & 0.39 & 16.6 &  2.3 & 15\\
$700 \leq P < 800$ days     & --4.64 & 0.50 & 20.5 &  2.7 & 4\\
$1440 \leq P \leq 2335$ days& --4.71 & 1.21 &  7.6 &  3.0 & 5\\
    \hline
$1000 \leq L < 1259$ L$_\odot$  & --8.61 & 0.32 & 15.0 &  4.0 & 3 \\
$1259 \leq L < 1585$ L$_\odot$  & --8.48 & 0.46 & 12.5 &  3.2 & 7$^\dag$ \\
$1585 \leq L < 1995$ L$_\odot$  & --8.47 & 1.31 & 11.9 &  3.2 & 5 \\
$1995 \leq L < 2511$ L$_\odot$  & --7.74 & 1.37 &  6.6 &  4.1 & 4 \\
$2511 \leq L < 3162$ L$_\odot$  & --6.39 & 0.48 &  7.8 &  4.3 & 5 \\
$3162 \leq L < 3981$ L$_\odot$  & --6.75 & 0.43 &  8.1 &  3.3 & 3 \\
$3981 \leq L < 5011$ L$_\odot$  & --6.47 & 0.71 &  9.9 &  5.1 & 19 \\
$5011 \leq L < 6309$ L$_\odot$  & --6.39 & 0.48 &  9.4 &  4.6 & 19 \\
$6309 \leq L < 7943$ L$_\odot$  & --5.92 & 0.69 & 13.7 &  4.1 & 20 \\
$7943 \leq L < 10000$ L$_\odot$ & --5.43 & 0.66 & 15.8 &  4.8 & 20 \\
$10000 \leq L < 12589$ L$_\odot$& --5.21 & 0.73 & 16.1 &  4.0 & 7 \\
$12589 \leq L < 15849$ L$_\odot$& --5.17 & 0.39 & 16.3 &  3.7 & 3 \\
$15849 \leq L < 10^5$ L$_\odot$ & --4.62 & 1.10 & 17.7 &  4.1 & 6 \\
$10^5 \leq L < 10^6$ L$_\odot$  & --3.53 & 0.96 & 20.7 &  5.2 & 2 \\
    \hline
\multicolumn{6}{p{0.47\textwidth}}{$^\ast$Excludes the super-giant AFGL 2343. $^\dag$Excludes the outlying NGC 2808 48889 from \citet{MCP06}.}\\
    \hline
\end{tabular}
\end{table}
\end{center}

\begin{center}
\begin{table}
\caption{Binned statistics of $K_{\rm s}-[22]$ colour at differing periods, for stars with $-1 < K_{\rm s}-[22] < 5$ mag.}
\label{K22BinTable}
\begin{tabular}{lcccc}
    \hline \hline
\multicolumn{1}{c}{Period} & \multicolumn{1}{c}{Mean}            & \multicolumn{1}{c}{Standard}  & \multicolumn{1}{c}{Fraction of}          & \multicolumn{1}{c}{Number}   \\
\multicolumn{1}{c}{range}  & \multicolumn{1}{c}{$K_{\rm s}-[22]$}& \multicolumn{1}{c}{deviation} & \multicolumn{1}{c}{stars with}           & \multicolumn{1}{c}{of} \\
\multicolumn{1}{c}{(days)} & \multicolumn{1}{c}{(mag)}           & \multicolumn{1}{c}{(mag)}     & \multicolumn{1}{c}{$K_{\rm s}-[22]>0.55$}& \multicolumn{1}{c}{stars} \\
    \hline
20--25 &   0.385 &0.461 &0.205 &  88\\
25--30 &   0.476 &0.584 &0.347 &  75\\
30--35 &   0.527 &0.671 &0.286 &  70\\
35--40 &   0.529 &0.749 &0.236 &  55\\
40--45 &   0.506 &0.409 &0.333 &  54\\
45--50 &   0.453 &0.440 &0.220 &  41\\
50--55 &   0.721 &0.639 &0.511 &  90\\
55--60 &   0.685 &0.616 &0.538 &  93\\
60--65 &   0.921 &0.577 &0.730 & 100\\
65--70 &   0.886 &0.588 &0.683 &  82\\
70--75 &   1.013 &0.572 &0.790 & 105\\
75--80 &   1.334 &0.803 &0.871 &  62\\
80--85 &   1.321 &0.786 &0.841 &  82\\
85--90 &   1.327 &0.549 &0.917 &  72\\
90--95 &   1.172 &0.729 &0.783 &  60\\
95--100 &  1.316 &0.660 &0.856 &  90\\
100--125 & 1.517 &0.668 &0.927 & 436\\
125--150 & 1.704 &0.650 &0.971 & 478\\
150--175 & 1.815 &0.652 &0.976 & 541\\
175--200 & 1.884 &0.648 &0.977 & 610\\
200--225 & 1.934 &0.542 &0.988 & 740\\
225--250 & 2.086 &0.606 &0.997 & 692\\
250--275 & 2.183 &0.638 &0.995 & 780\\
275--300 & 2.296 &0.629 &0.999 & 749\\
300--400 & 2.423 &0.781 &0.991 &1704\\
    \hline
\end{tabular}
\end{table}
\end{center}

To determine the physical and observable criteria for the onset of dust production in solar-neighbourhood stars, we now examine the increase in mass loss over the $\sim$60-day period boundary. Note that this is not a hard boundary, but one that occurs between periods of $\sim$50 and $\sim$100 days, depending on the star in question.

Figure \ref{ColDensFig} shows a clear link between the infrared colour of the star and its mass-loss rate across the $\dot{M} \sim 10^{-7}$ M$_\odot$ yr$^{-1}$ regime. The slope of the ($K_{\rm s}-[22]$)--$\log(\dot{M})$ relation of Equation \ref{K22MdotEqns} (0.464; Figure \ref{ColDensFig}; Section \ref{EmpiricalSect}) is sufficiently close to 0.4 to indicate that the dust condensation efficiency remains approximately constant across the measurable regime, until $K_{\rm s} - [22] \approx$ 5 mag, where the dust becomes optically thick at $K_{\rm s}$ and the star becomes optically invisible.  The scatter (standard deviation) away from the relation is 0.75 dex. Hence, we can establish that $K_{\rm s}-[22]$ is a good tracer of stellar mass-loss rate to better than a factor of six, as averaged over our dataset. We may therefore expect from this alone that the increase in $K_{\rm s}$--[22] colour at $P \sim 60$ days is a real increase in mass loss, not just an increase in dust condensation.

Figure \ref{MLRVexpFig} shows that the mass-loss rate is tightly correlated with pulsation period, but that stars with a given period and mass-loss rate scatter in luminosity (e.g., the outlier $\alpha$ Her). Quantifying this is difficult: although we have a well-spread range of stars in luminosity, there are no stars in the ranges $30 < P < 40$ days, $46 < P < 55$ days, $235 < P < 300$ days or $795 < P < 1440$ days. Table \ref{BinTable} lists the mass-loss rates for various different ranges of period and luminosity. Among the measured stars, mass-loss rates for stars with $P \leq$ 46 days are consistently low (a few $\times 10^{-9}$ M$_\odot$ yr$^{-1}$). However, stars with periods $55 \leq P < 400$ days closely approximate a constant mass-loss rate of $\dot{M} \approx 3.7 \times 10^{-7}$ M$_\odot$ yr$^{-1}$, with a scatter of only a factor of three. Mass-loss rates then rise rapidly with period, so that by $\sim$700 days they tentatively appear to stabilise near $\dot{M} \sim 2 \times 10^{-5}$ M$_\odot$ yr$^{-1}$.

By contrast, the average $\dot{M}$ with luminosity typically exhibits greater scatter. While we caution that the number of stars in each bin is low, between the RGB tip and $\sim$6300 L$_\odot$ (where the increase displayed by \citet{DTJ+15} begins; see also Figure \ref{MLRVexpFig}) there is a scatter of factors of 2.5 to 20 in a given luminosity bin. This is in contrast to the roughly constant factor of 2.3 to 4.0 seen in the corresponding period bins.

The outflow velocities (also in Table \ref{BinTable}) are more difficult to interpret. There is no clear link between period and $v_{\rm exp}$ until $P \sim 400$ days, then a clear increase to $P \sim 800$ days. Stars with $P \geq 1440$ days do not follow the above trends: these are typically supergiant stars ($\alpha$ Ori, OH 104.9+2.4, OH 26.5+0.6, V669 Cas and AFGL 5379). In comparison, there appears a declining outflow velocity with luminosity until the RGB tip ($\sim$2200 L$_\odot$), beyond which it stabilities at minimum until $L \sim 6300$ L$_\odot$. There is then a notable rise of outflow velocity with luminosity that continues to the brightest supergiants. The scatter in each period bin ($\sim$3 km s$^{-1}$) is typically lower than in each luminosity bin ($\sim$4 km s$^{-1}$); however, the biases in our sample prevent us from definitively stating that this is significant.

We are therefore left with the conclusion that mass-loss rate correlates considerably better with pulsation period than luminosity, and that a roughly constant mass-loss rate occurs between $P \sim 55$ and $\sim$400 days. Pulsation period therefore appears to define the mass-loss rate of AGB stars (though not supergiants) up to \emph{at least} $P \sim 400$ days and $L \sim 6300$ L$_\odot$. Links to the outflow velocity are less clear, tracking both pulsation and luminosity, though a minimum velocity of $\sim$8 km s$^{-1}$ may occur somewhere near the RGB tip.

\subsubsection{The role of pulsation amplitude}
\label{PulsAmpSect}

While pulsation is typically measured as a $V$-band amplitude, optical changes in flux result mostly from changes in molecular opacity, and pulsation amplitude is better measured on the Rayleigh--Jeans tail, where radial effects dominate and temperature-sensitive molecules (like TiO) are largely transparent \citep[e.g][]{BHN+13}. Consequently, differences between Miras ($\Delta V > 2.5$ mag) and periodic semi-regular variables ($\Delta V < 2.5$ mag) may not be meaningful. We adopt the $K$-band amplitudes of \citet{PSK+10} provide a better measure of pulsation strength, and indeed Figure \ref{KAmpMLRFig} shows that $V$-band and $K$-band amplitude are poorly correlated.

Stars with $P \lesssim 60$ days are universally not identified as variable in \citet{PSK+10} and have visual (GCVS) amplitudes of $\Delta V \leq 1.11$ mag ($\Delta V \leq 0.63$ mag below 55 days). Stars with $60 \lesssim P \lesssim 300$ days are included frequently included in \citet{PSK+10}, with $K$-band amplitudes of typically a few per cent, and visual amplitudes of $\Delta V = 0.2$--4.4 mag. However, there is no clear link between $K$-band amplitude and mass-loss rate until $\Delta F_K / F_K \sim 0.3$, which is achieved at $P \sim 400$ days, whereupon mass-loss rate increases by roughly an order of magnitude (Figure \ref{KAmpMLRFig}). This may simply be due to the small number of stars in the range $60 \lesssim P \lesssim 300$ with good observations, or it may reflect that pulsation amplitude is not the primary deciding factor of mass-loss rate.

\subsubsection{The link to $K_{\rm s}-[22]$ colour and radiation pressure on dust}

Table \ref{K22BinTable} similarly details the binned $K_{\rm s}-[22]$ colour versus period for the entire GCVS sample used in \citet{MZ16}. In practice, this sample will only contain high-amplitude variables. Therefore, if pulsation amplitude is related to mass-loss rate (Section \ref{PulsAmpSect}), then this will likely represent an upper limit to the average values.

The average $K_{\rm s}-[22]$ colour remains roughly constant at $\sim$0.5 mag, up to $P \sim 50$ days, with approximately a quarter of recorded variable stars showing infrared excess ($K_{\rm s}-[22] > 0.55$ mag. Both quantities then rapidly increase such that, by $P \sim 80$ days, the average $K_{\rm s}-[22] \approx 1.3$ mag and 80--90 per cent of stars show excess. The increase flattens off, with the quantities being $\approx$1.8 mag and 97 per cent by 150 days. However, both the average colour and percentage of stars with infrared excess continue to slowly increase with stellar period, reaching $\sim$2.5 mag and unity by $P \sim 400$ days: this is in contrast to the measured CO mass-loss rate and wind velocity.

The uncertainties in the average mass-loss rates over this regime (150--400 days) are still considerable (Table \ref{BinTable}). Hence, without further observations of stars in the $P \sim 150$ day range, it is impossible to conclusively determine whether this increase in $K_{\rm s}-[22]$ colour by 0.7 mag represents a real increase in mass-loss rate by a factor of two, or an increase in dust opacity per unit mass in longer-period stars. One possibility is that extra opacity is needed to overcome the stellar gravity in long-period stars, which originate from higher-mass progenitors, so are more likely to be massive and denser during the AGB phase than their shorter-period counterparts.


\subsection{Limitations and biases in the approach}
\label{BiasSect}

For a number of reasons, a more robust and statistical treatment of these data is not practical at present. This is due to a number of limitations of the data, which first need to be addressed by the community at large.

First and foremost, strong observational biases exist in our data (historically, there is a preference for performing CO observations of apparently bright and dusty AGB stars). Hence, the statistical averages presented in this work may not be wholly representative of the underlying population. Nevertheless, the relative statistics within the dataset are expected to broadly reflect any physical changes. A volume-limited sample of AGB stars is needed to properly address these issues.

Distances to AGB stars suffer uncertainties due to lack of direct measurement, resulting in large uncertainties in the luminosities and radii of stars. In the majority of literature cases, stars are assumed to be fundamental-mode pulsators, and a luminosity is derived from a period--luminosity relation. In many cases, luminosities can only be determined to within a factor of two. If mass-loss rate or outflow velocity scales better with pulsation period than with luminosity, this may cause a tighter relationship to be seen in Figure \ref{MLRVexpFig} than is real; or it may cause additional scatter if the reverse is true.

Direct distance measurements are often limited to optical parallaxes. Two widely-acknowledged but significant factors affect the accuracy of parallax distances of long-period variables. Firstly, in cool giants, the motion of the centre of light across the stellar surface creates a long-term wander of the astrometric centroid \citep[e.g.][]{vanLeeuwen07}. If this shift correlates with the annual parallax effect, it can create an unanticipated change (usually an increase) in the stellar parallax. The effect should be most prominent in the physically largest and coolest stars. Secondly, treatment of outliers in the \emph{Gaia} data mean that observations taken at photometric extrema of large-amplitude variables may be discarded from the astrometric solution. If this selection effect correlates with the observing cadence, it can lead to errors in the parallax computation\footnote{The \emph{Gaia} Data Release highlights the case of RR Lyrae, which is assigned a negative parallax for this reason.}. The effect should be largest in the most variable stars. It is expected that further these problems will be largely resolved as the \emph{Gaia} satellite collects more data, and data reduction techniques become more robust for strongly coloured and highly variable sources.

Methods and definitions for effective temperature and radius of AGB stars also need homogenising. An effective temperature scale relies on being able to define a radius in order to invoke $L \propto R^2 T_{\rm eff}^4$. However, the photospheric radius for AGB stars is highly wavelength-dependent due to a combination of molecular effects and circumstellar dust, meaning the `surface' of the star becomes a semantic problem. Furthermore, stars are highly non-uniform, and optical tracers of temperature only represent the output from the hottest parts of the stellar surface \citep[e.g.][]{FLH17}. Stars are also highly variable, meaning that observing or defining a fixed spectrum for calculations is extremely difficult. The presence of circumstellar absorption by dust and re-emission of infrared light makes comparison to theoretical spectra difficult, and photometric methods impractical (e.g., the infrared flux method; \citet{BS77}). Consequently, temperatures derived from spectroscopy suffer from lack of calibration, difficulties in reproducing molecular opacity in stellar models, difficulties in determining stellar metallicity, stellar surface inhomogeneities and effects caused by being out of local thermodynamic equilibrium, so that accurately defining a temperature and radius is extremely difficult \citep[e.g.][]{LHA+12}. For dust-producing stars, temperatures derived from photometry are skewed towards cooler temperatures due to circumstellar reddening by dust in the optical, and emission from that dust in the infrared. Modelling of this dust can `correct' the effective temperature to approximate that of the near-infrared photosphere, but often with a strong degeneracy between dust opacity and stellar temperature. Consequently, spectroscopic and photometric identifications of temperature can differ significantly, even for very well-characterised stars (cf., \citealt{LNH+14} versus \citealt{MBvLZ11}). Consequently, we should not expect to see good correlation between mass-loss rate or expansion velocity, and effective temperature or radius for the highest mass-loss rate AGB stars: even if one were to exist, these terms lack coherent definition across the literature of AGB stars. We therefore have herein reduced our comparisons to observational correlations with inferred luminosity and pulsation properties, whereas radius and temperature may be important parameters on which the wind depends.

Proxy measurements of mass-loss rate and expansion velocity from stellar chromospheres need better calibration against CO-based mass-loss rates. These proxies rely on measuring gas motions in the stellar chromosphere, very close to the surface, where material is still bound to the star and not yet subject to the full acceleration of radiation pressure on dust. Rates and velocities derived from chromospheric outflows are also highly time-variable \citep[e.g][]{MvLDB10}, recording a more instantaneous measure of mass loss than CO rotational lines. Consequently, it is not clear how well an individual measurement represents the time average of the wind properties, nor that material passing through the chromosphere will actually be lost from the star, rather than falling back. Typically H$\alpha$ and Ca {\sc ii} lines are used for this work, so they have the advantage of tracing atomic and low-ionisation conditions, but UV lines can also be used for highly ionised winds. Addressing this potential source of bias will require measurement of mass-loss rate from a diverse range of well-studied stars with a number of techniques.

\subsection{Future expectations and remaining questions}

Several questions remain in this analysis, and several statements are conjectural at present. These include:

{\it Physically, how do pulsations dictate the mass-loss rate?} Pulsations levitate material from the star, setting the initial conditions in the dust-formation zone. However, models of pulsating stars \citep[e.g.][]{BHAE15} do not cover the overtone pulsators we observe. Further exploration of this parameter regime would be useful to know whether existing models can drive a wind under these circumstances.

{\it What role does stellar mass/gravity have in tempering the mass-loss rate and wind velocity?} Stellar period and amplitude appear to largely dictate mass-loss rate, even for fairly massive stars like $\alpha$ Her ($2.175 \lesssim M \lesssim 3.25$ M$_\odot$; \citealt{MGKW13}). However, a statistical look at a large sample of stars with known masses would be necessary to have a more consistent view of the effects of mass. This may explain differences in the mass loss from supergiants like $\alpha$ Ori \citep[e.g.][]{RDD+13} from stars like VY CMa \citep[e.g.][]{RIH+14,OVR+15}, and relate them to their lower-mass counterparts.

{\it How do these parameters change with stellar metallicity and environment?} While chromospheric winds appear to have little metallicity independence \citep{MZ15b}, the amount of condensible material that can form dust is less in metal-poor stars. Yet metal-poor stars are prodigious dust producers \citep{BMB+15,MvLS+11}, even though the constituents of that dust may be different \citep{MSZ+10,JKS+12}. In the Magellanic Clouds, there exists a population of `anomalous' stars with similar infrared colours to our 60--300-day Galactic sample \citep{BMS+15}, and super-solar-metallicity AGB stars show no extra mass-loss either \citep{vLBM08}. Consequently, we may expect there to be very little metallicity dependence in the (gas) mass-loss rates from stars, while existing observations in the Magellanic Clouds indicate at least outflow velocity may decline at low metallicity, indicating metallicity may have an effect on the outflow velocity instead \citep{GVM+16,MSS+16,GvLZ+17}. A survey of mass-loss rates from truly metal-poor stars over a range of luminosities is needed. This has been argued for since \citet{Reimers75}, but is only now being realised \citep[e.g][]{BMB+15,BMG+17}.

{\it What is the role of stellar chemistry?} Since carbon stars produce their own dust-forming materials, we may expect them to exhibit different characteristics \citep[cf.][]{LZ08}. Yet no differences of note are visible in the plots shown above. It may be in metal-poor stars that we start to see a divergence in the mass-loss rates between carbon stars (which remain effective mass losers) and oxygen-rich stars (which lack dust to drive their winds).

{\it How are these changes reflected in the dust properties of the star, and its mineralogical return to the interstellar medium?} As well as changes in opacity between dust from metal-rich and metal-poor stars, and between carbon- and oxygen-rich dust, there is a wide variety of dust composition in AGB stars of a particular metallicity and chemical type \citep[e.g.][]{WOK+11}. The causes of this variation in mineralogy require explanation in the context of the mass-loss process.

{\it What is magical about the periods of 60 and 300 days?} In \citet{MZ16}, we argued that the 60-day period represents the point at which the lowest-mass stars become first-overtone pulsators, and that their pulsations levitate material so that dust can form. However, the role of the 300-day period is less clear. Stars transition to the fundamental mode at periods much lower than 300 days \citep[e.g.][]{Wood15}. We suggest that this may be when a radiation-driven wind becomes effective, yet it is not obvious what is special about this period in particular.

{\it What is the role of binarity and asymmetries?} We have assumed here that stars are isolated and spherical. Yet we know that environment alters both the mass-loss rate of stars and how we measure that mass-loss rate, due to the chemical and morphological factors involved. Problems include variation in radiation environment \citep[e.g.][]{MGH88,MZ15a,MZ15b}, shaping of ejecta \citep[e.g.][]{MMV+12,MRLF+14,DRN+15,RMV+14} and the formation of circumbinary discs \citep[e.g.][]{LKP+15,KHR+16,HRD+17}. A statistical study of binary AGB stars is needed to understand these effects in the context of the evolution of whole stellar populations.


\section{Conclusions}
\label{ConcSect}

Based on new and literature observations of mass-losing AGB stars, we provide strong evidence that pulsations are the primary mechanism setting the mass-loss rates of most AGB stars. The mass-loss rate and expansion velocity of AGB-star winds is found to correlate poorly with stellar luminosity, contrary to the expectations of a radiation-driven wind, but strongly with stellar pulsation period and near-infrared amplitude. Radiation pressure on dust appears to play a subsidiary role in the mass-loss process, likely setting factors like grain size and outflow velocity, as required to maintain the mass-loss rate dictated by the pulsations.

We calculate that $\dot{M} \lesssim 10^{-8}$ M$_\odot$ yr$^{-1}$ among stars with $P \lesssim 55$ days. This rises to an average of $\dot{M} \approx 3.7 \times 10^{-7}$ M$_\odot$ yr$^{-1}$ among measured stars, over the regime $55 \lesssim P \lesssim 400$ days. A further, well-documented increase extends to $\dot{M} \sim 3 \times 10^{-5}$ M$_\odot$ yr$^{-1}$ by $P \approx 700$ days, beyond which stars become few in number and disparate in their properties. The precise boundaries of these transitions remain unclear, however, and the mass-loss rates listed here may be heavily biased by selective observations of past observers. Nevertheless, the increase of a factor $\sim$100 in mass-loss rate near a period of 60 days suggest existing mass-loss prescriptions in stellar evolution models need rethought and recalibrated, both in terms of RGB and AGB mass loss. This work can be thought of as a first step in that process.

Within these rates, individual stars vary by significant factors. Further observations are recommended of stars with $P \lesssim 300$ days, and more accurate modelling of existing observations encouraged, to better constrain their physical parameters. Unlike radiation-driven winds, we predict that the above period dependences of mass-loss rate are mostly independent of stellar metallicity, but may be strongly dependent on stellar mass. We strongly encourage further, unbiased CO observations of stars with $P < 300$ days, and modelling to test these hypotheses.


\section*{Acknowledgements}

Based on observations with the Atacama Pathfinder EXperiment (APEX) telescope. APEX is a collaboration between the Max Planck Institute for radio Astronomy, the European Southern Observatory, and the Onsala Space Observatory. The authors acknowledge support from the UK Science and Technology Facility Council under grant ST/L000768/1. IRAF is distributed by the National Optical Astronomy Observatory, which is operated by the Association of Universities for Research in Astronomy, Inc., under co-operative agreement with the National Science Foundation.


\label{lastpage}

\end{document}